\documentclass[aps,pra,twocolumn,superscriptaddress,eqsecnum,showpacs]{revtex4} 
\usepackage{graphicx,amssymb,amsmath,bm,times}
\allowdisplaybreaks

\begin{document}

\title{Composite pairing and superfluidity in a one-dimensional 
resonant Bose-Fermi mixture\! }

\author{Shimul Akhanjee}
\affiliation{Condensed Matter Theory Laboratory, RIKEN, 
Wako, Saitama, 351-0198, Japan}
\affiliation{Department of Condensed Matter Physics and Materials Science, 
Brookhaven National Laboratory, Upton, New York 11973, USA}

\author{Masahisa Tsuchiizu}
\affiliation{Department of Physics, Nagoya University, Nagoya 464-8602, Japan}

\author{Akira Furusaki}
\affiliation{Condensed Matter Theory Laboratory, RIKEN, 
Wako, Saitama, 351-0198, Japan}
\affiliation{RIKEN Center for Emergent Matter Science (CEMS), 
Wako, Saitama, 351-0198, Japan}

\date{October 17, 2013}

\begin{abstract}
We study the ground-state properties of one-dimensional mixtures of
bosonic and fermionic atoms resonantly coupled to fermionic Feshbach
molecules. 
When the particle densities of
fermionic atoms and Feshbach molecules differ, the
system undergoes various depletion transitions between binary and
ternary mixtures, as a function of the detuning parameter. However,
when the particle densities of fermionic atoms and Feshbach molecules
are identical, the molecular conversion and disassociation processes induce a
gap in a sector of low-energy excitations, and the remaining system
can be described by a two-component Tomonaga-Luttinger liquid.
Using a bosonization scheme, we derive the effective low-energy
Hamiltonian for the system, which has a similar form as that of the
two-chain problem of coupled Tomonaga-Luttinger liquids.  With the help
of improved perturbative renormalization group analysis of the latter
problem, we determine the ground-state phase diagram and find that
it contains a phase dominated by composite
superfluid or pairing correlations between the open and closed
resonant channels.
\end{abstract}


\maketitle

\section{Introduction}
\vspace*{-.3cm}

The Feshbach resonance \cite{feshbach}, as experimentally realized in
ultracold atoms and molecules in optical lattices, has made it possible
to investigate the many-body physics of multicomponent quantum
degenerate mixtures of fermions and/or bosons with interspecies interactions
\cite{jinPRL2004,ketterlePRL2004,ThomasPRL2004,chinSCIENCE2004,blochrmp08}.
Operationally, a magnetic field near resonance can tune the energy
splitting between different hyperfine configurations of atoms,
yielding a tunable scattering amplitude with a magnitude that depends
on the mismatch of the magnetic moments \cite{pethickbook}.  In this
context, theoretical studies have introduced two primary interaction
vertices: a short-ranged, one-channel density-density type
interaction and a two-channel interaction that couples open-channel
atoms to a molecular bound-state (MB) particle
\cite{Kokkelmans:2002gm,Bruun:2004kz,Dulieu:2009ei}.

Recently, heteronuclear fermionic Feshbach molecules composed of
bosonic ${}^{23}$Na and fermionic ${}^{6}$Li \cite{Stan:2004ic} and
of bosonic ${}^{87}$Rb and fermionic ${}^{40}$K \cite{Inouye:2004en}
have been observed experimentally and attracted the attention of
theoretical studies
\cite{Yabu2003,Yabu:2004ha,adhikariPRA04,powellPRB05,Dickerscheid:2005ey,Zhang:2005dr,Avdeenkov:2006ef,Bortolotti:2008in,bfrpa}
focusing on the competition between the condensed state of unpaired
bosons and the degenerate MB particles with an
additional Fermi surface.  It has been argued that there can be
depletion transitions \cite{Yabu2003,Yabu:2004ha,powellPRB05} where
one or more of the atomic or molecular species can be exhausted by
driving the formation or disassociation of MB particles.  Furthermore,
if bosons are condensed, the spectrum can be directly diagonalized,
yielding MB particles that are dressed by free atomic fermions, which
form low-energy quasiparticles in a Fermi-liquid theory \cite{bfrpa}.
Additionally, the superfluidity of a paired state of a fermionic
atom and a fermionic molecule, which is formed through attractive
interactions mediated by the condensed and/or uncondensed bosons, has
been predicted to occur \cite{Zhang:2005dr}.  However, it is
questionable as to whether such features obtained by a mean-field
approach can persist when strong quantum fluctuations are present,
especially for atoms trapped in one-dimensional (1D) tubes.

There are many reliable analytical and numerical methods available for
1D systems \cite{Cazalilla:2004dd,Cazalilla_review}.  In particular,
the bosonization technique has been applied to one-channel systems
with density-density type interactions, showing pairing and
density-wave instabilities \cite{cazalillaPRL03}, polaronic phases
\cite{matheyPRL2004,matheyPRA2007,Danshita:2013hw}, and competing orders
\cite{tsuchiizuPRA2010}.  The dominant phases exhibit variants of
``paired'' order parameters with algebraic decay or quasi-long-range
order (QLRO) \cite{cazalillaPRL03,matheyPRL2004,matheyPRA2007,Danshita:2013hw}.
Systematic analysis has also been performed for a two-channel type
model arising from atom-molecule mixture, expected for narrow
resonances \cite{Sheehy2005,Orignac:2006cb}; however, these investigations were
primarily focused on the bosonic MB particles for two-component fermions
in the context of the BEC-BCS crossover. 
 The possibility of more complex pairing
and superfluid orders that couple the open and closed fermionic
channels has not been observed experimentally or discussed
theoretically in detail.

In this paper, we study a general two-channel model of fermionic and
bosonic atoms near a narrow Feshbach resonance where bosons, fermions,
and molecules can coexist.  Using a renormalization-group (RG) method
based on the bosonization formalism, we obtain a low-energy theory
and attempt to clarify the ground-state phase diagram, with an emphasis
on the conditions that allow the pairing of the fermionic atoms and
molecules across the Feshbach resonance.  In doing so, we make use of
the analogy to the two-chain problem of coupled Tomonaga-Luttinger liquids
(TLL).  
The paper is organized as follows.
In Sec.\ \ref{sec:2}, we introduce the model and examine the
condition for ternary mixed phases of bosonic atoms, fermionic atoms,
and fermionic molecules.  In Sec.\ \ref{sec:3}, 
the ternary mixed phase is studied
and possible order parameters are
introduced to characterize QLRO.  We determine the phase diagram for
the case of an incommensurate density regime of fermions and molecules.
In Sec.\ \ref{sec:RG}, the RG method is applied to analyze
the low-energy properties, and in Sec.\ \ref{sec:5}, the phase diagram is
determined for the commensurate density regime of fermions and
molecules.  
Lastl, in Sec.\ \ref{sec:6} we summarize our results in the conclusion.
It so happens that, given the mathematical form of the resonant interaction, we can draw on an RG approach applied to the spinless
two-coupled chain, which is revisited in Appendix \ref{sec:app-A}.
Finally, as a supplement, we present an alternative approach based on a gauge transformation procedure in
Appendix \ref{sec:app-B}.

\vspace*{-.3cm}

\section{Model and condition for ternary mixed phase}\label{sec:2}

\vspace*{-.3cm}

\subsection{Model Hamiltonian}

\vspace*{-.3cm}

Our starting point is a coupled, two-channel model that describes a resonant
scattering process, where free bosonic ($b$) and fermionic ($f$) atoms resonate into 
fermionic Feshbach molecules ($\psi$).
The model Hamiltonian is given by
\begin{equation}
H=
H_b+H_f+H_{\psi}+H_{3p},
\label{eq:fullham}
\end{equation}
where
\begin{subequations}
\begin{eqnarray}
H_b 
\!\! &=& \!\!\!
\int dx \,
  \Psi_b^\dag(x)
\left(
- \frac{1}{2m_b}\frac{d^2}{dx^2} - \mu_b
\right)
\Psi_b(x) 
\nonumber\\
&& {} 
+ \frac{1}{2} \int dx dx' V_{bb}(x-x') \rho_b(x) \rho_b(x') ,
\\
H_f 
\!\! &=& \!\!\!
\int dx \,
  \Psi_f^\dag(x)
\left(
- \frac{1}{2m_f}\frac{d^2}{dx^2} - \mu_f
\right)
\Psi_f(x) 
\nonumber\\
&& {} 
+ \frac{1}{2} \int dx dx' V_{ff}(x-x') \rho_f(x) \rho_f(x') ,
\\
H_{\psi} 
\!\! &=& \!\!\!
\int dx  \,
  \Psi_\psi^\dag(x)
\left(
- \frac{1}{2m_\psi}\frac{d^2}{dx^2} + \nu - \mu_\psi
\right)
\Psi_\psi(x) 
\nonumber\\
&& {} 
+ \frac{1}{2} \int dx dx' V_{\psi\psi}(x-x') \rho_\psi(x) \rho_\psi(x')
,
\qquad
\label{eq:hamilt-psi}
\\
H_{3p} 
\!\! &=& \!\!
g_{3p}
\int {dx\left[ {\Psi_\psi^\dag(x) \Psi_f(x) \Psi_b(x) +
 \mathrm{H.c.}
}
	      \right]},
\label{eq:hamilt-g3}%
\end{eqnarray}%
\label{eq:hamilt}%
\end{subequations}
and we have set $\hbar=1$.  The density operators are
$\rho_s(x)=\Psi_s^\dagger(x) \Psi_s(x)$, ($s=b,f,\psi$), where the
field operators $\Psi_s(x)$ obey the usual commutation and
anticommutation relations for bosons ($s=b$) and fermions
($s=f,\psi$).  The Hamiltonian $H_s$ ($s=b,f,\psi$) consists of a
kinetic energy term and an intraspecies density-density interaction
term.  The coupling $g_{3p}$ in Eq.\ (\ref{eq:hamilt-g3}) induces the
conversion of bosonic ($b$) and fermionic ($f$) atoms into fermionic
MB particles ($\psi$) and vice versa (disassociation)
\cite{Kokkelmans:2002gm,Bruun:2004kz}.  The individual
particle numbers are not conserved; instead, the total numbers of
bosonic and fermionic atoms,
\begin{subequations}
\begin{eqnarray}
\mathcal{N}_B
\!\! &=& \!\!\!
\int dx
\left[
\rho_b(x) + \rho_\psi(x) 
\right] ,
\\
\mathcal{N}_F
\!\! &=& \!\!\!
\int dx
\left[
\rho_f(x) + \rho_\psi(x) 
\right] ,
\end{eqnarray}%
\label{eq:N}%
\end{subequations}
are conserved quantities.  It follows that the masses ($m_s$) and the
chemical potentials ($\mu_s$) obey the sum rules for mass conservation
and chemical equilibrium,
\begin{equation}
m_b+m_f=m_\psi , \quad
\mu_b+\mu_f=\mu_\psi ,
\label{eq:eqmu}
\end{equation}
and the detuning parameter $\nu$ in Eq.\ (\ref{eq:hamilt-psi})
defines the energy splitting between
the open and closed channels. 
The fermionic
intraspecies couplings, $V_{ff}(x)$, and $V_{\psi\psi}(x)$ are assumed to be
short-ranged, while the $b$ atoms interact with each other through the coupling
$V_{bb}(x)$.
At strong repulsion, the
boson system is described by an ordinary Tonks-Girardeau (TG) gas which
behaves as free fermions.

\begin{figure}[b]
\includegraphics[width=8cm]{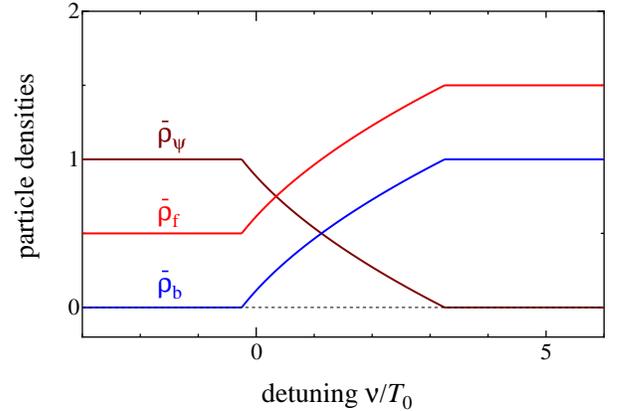}
\caption{
(Color online)
The normalized particle densities $\bar{\rho}_s \equiv
 L\rho^0_s/\mathcal{N}_B$ as a function of the detuning 
parameter $\nu$.
In this figure, we choose $m_b=m_f$ and $\mathcal N_F/\mathcal N_B=3/2$.
At $\nu/T_0 = 11/32$, the densities of fermions and molecules
become equal, $\bar{\rho}_f = \bar{\rho}_\psi = 3/4$.
}
\label{fig:density-limit}
\end{figure}

\vspace*{-.3cm}

\subsection{Phase diagram in the limit of $g_{3p}\to 0$}

\vspace*{-.3cm}

Before proceeding to the many-body features of the model described by
Eq.\ (\ref{eq:fullham}), it is important to first establish the range of
physical parameters that allow the ternary coexistence of all atoms and
molecules.
For simplicity, we will consider the limit $g_{3p}\to 0$, with
Tonks-Girardeau bosons [$V_{bb}(x)=g_b \delta (x)$ with $g_b\to + \infty$],
and noninteracting fermions and molecules [$V_{ff}(x)=V_{\psi\psi}(x)=0$].
As noted in Ref.\ \cite{powellPRB05},
we can construct a set of dimensionless parameters
$\mathcal{N}_F/\mathcal{N}_B$, $m_f/m_b$, and $\nu/T_0$, where 
$T_0$ is the ``Fermi'' degeneracy temperature for hard-core bosons:
$T_0\equiv \pi^2 \mathcal{N}_B^2/(2m_b L^2)$, with $L$ being the system size.

Let us introduce the average particle density 
$\rho^0_s=L^{-1}\int \rho_s(x)dx$
and the corresponding normalized quantity
$\bar\rho_s\equiv L\rho_s^0/\mathcal{N}_B$.
The conditions for
the conserved total numbers of atoms [Eqs.\ (\ref{eq:N})] are expressed as
$1=\bar\rho_b+\bar\rho_\psi$ and
$\mathcal{N}_F/\mathcal{N}_B=\bar\rho_f+\bar\rho_\psi$,
respectively.
For hard core bosons, free fermions, and free molecules,
the chemical potentials are given by
 $\mu_b=(k_F^b)^2/(2m_b)$, 
 $\mu_f=(k_F^f)^2/(2m_f)$, and
 $\mu_\psi=(k_F^\psi)^2/(2m_\psi) + \nu$,
where 
the ``Fermi momenta'' for each species are given by
\begin{equation}
k_F^s=\pi \rho^0_s.
\end{equation}

\begin{figure}[t]
\includegraphics[width=8cm]{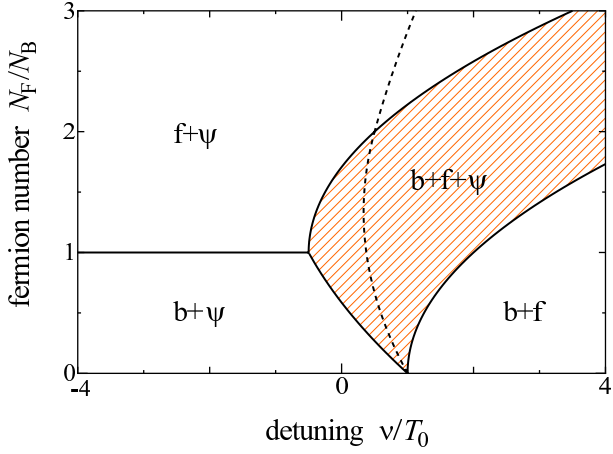}
\caption{
(Color online)
Phase diagram in terms of the detuning parameter $\nu$ and the 
 fermion number $\mathcal N_F$
for the case of equal masses $m_b=m_f$.
The ternary mixed state of bosonic atoms, fermionic atoms, 
and Feshbach molecules is realized in  
 the region denoted by ``$b$+$f$+$\psi$.''
The regions denoted by ``$f$+$\psi$,'' ``$b$+$\psi$,'' and ``$b$+$f$'' 
represent the fermion-molecule, boson-molecule, and boson-fermion
binary mixed phases, respectively.
Along the dashed line,
the densities of fermions and molecules become 
equal, $\bar \rho_f=\bar
 \rho_\psi$.
}
\label{fig:pd-limit}
\end{figure}

The particle densities can be determined from 
the equilibrium condition of Eq.\ (\ref{eq:eqmu}).
In the ternary mixed phase of 
$b$, $f$, and $\psi$ particles ($b$+$f$+$\psi$ phase), 
the density of molecules $\bar\rho_\psi$ is determined by the
following equation:
\begin{equation}
\left(1-\bar{\rho}_\psi\right)^2 
+ \frac{1}{\bar{m}_f} \left(\bar{\mathcal N}_F-\bar{\rho}_\psi\right)^2 
=
 \frac{1}{1+\bar{m}_f} \left(\bar{\rho}_\psi\right)^2 + \bar{\nu},
\end{equation}
where $\bar{\mathcal N}_F\equiv \mathcal N_F/\mathcal N_B$,
$\bar{m}_f\equiv m_f/m_b$, and $\bar\nu \equiv \nu/T_0$.
The densities for $b$ atoms and $f$ atoms are determined by 
$\bar{\rho}_b=1-\bar{\rho}_\psi$ and
$\bar{\rho}_f=\bar{\mathcal N}_F-\bar{\rho}_\psi$, respectively,
and the expected $\nu$ dependence is shown in Fig.\ \ref{fig:density-limit}.
Notice that in the case of sufficiently strong positive detuning, 
the $\psi$ particle is completely depleted and only the $b$ and $f$
atoms remain.
We thus label this binary mixture the ``$b$+$f$'' phase;
effects of possible heteroatomic interactions 
in this regime have been analyzed in the literature 
\cite{dasPRL2003,cazalillaPRL03,matheyPRL2004,%
matheyPRA2007,Danshita:2013hw,polletPRL2006,tsuchiizuPRA2010}, 
where it has been pointed out 
that the excitation spectrum can have a gap and the pairing fluctuations 
are enhanced when the particle densities of two kinds of atoms become
equal. 
On the other hand, 
the MB particles become stable for sufficiently strong negative detuning;
for $\mathcal N_F/\mathcal N_B < 1$ $(>1)$,
either $b$ or $f$ atoms coexist with the $\psi$ particles
and the resulting binary mixtures are labeled
``$b$+$\psi$ '' and ``$f$+$\psi$ '' phases, respectively.

The phase diagram in terms of the detuning parameter $\nu$ and the
total fermion number $\mathcal N_F$ is shown in Fig.~\ref{fig:pd-limit},
which can be contrasted with the corresponding
phase diagram in the
three-dimensional (3D) case (see Fig.\ 3 in Ref.\ \cite{Yabu2003},
Fig.~3 in Ref.\ \cite{powellPRB05}, and also Fig.~1 in
Ref.\ \cite{Zhang:2005dr}), where the
Bose-Einstein condensate (BEC) proliferates everywhere except for the 
small $\nu$ and large $\mathcal{N}_F$ region
 corresponding to the $f$+$\psi$ phase in Fig.\ \ref{fig:pd-limit}.
In the present 1D case, no BEC can occur in any parameter region, but
a ``Fermi surface'' of the $b$ atoms can be observed instead.
With this in mind, we find qualitative agreement 
with our phase diagram for 1D mixtures and that for 3D mixtures. 
The densities of fermions and molecules become identical 
($\bar{\rho}_f = \bar{\rho}_\psi$) in both ternary and binary 
mixed phases for a particular $\bar{\nu}$, satisfying
\begin{equation}
\bar{\nu}
=
\left\{
\begin{array}{lll}
\displaystyle
1 - \bar{\mathcal N}_F
+ \frac{1+\bar m_f+ \bar m_f^2}{4\bar m_f(1+\bar m_f)} 
\bar{\mathcal N}_F^2
&&
\mbox{($b$+$f$+$\psi$ phase)},
\\ \\
\displaystyle
\frac{1}{4} 
\left(\frac{1}{\bar{m}_f}- \frac{1}{1+\bar{m}_f}\right) 
\bar{\mathcal N}_F^2
&& 
\mbox{($f$+$\psi$ phase)},
\end{array}
\right. 
\end{equation}
which is represented by the dashed line in Fig.\ \ref{fig:pd-limit}.
The analysis given in Ref.\ \cite{cazalillaPRL03} may be applied
to the case $\bar\rho_f=\bar\rho_\psi$ in the $f$+$\psi$
phase. However, the spectrum for the case $\bar\rho_f=\bar\rho_\psi$
in the $b$+$f$+$\psi$ phase has not yet been analyzed so far.
In the following sections, we study phases realized inside the
$b$+$f$+$\psi$ phase upon turning on the $g_{3p}$ coupling.

\vspace*{-.3cm}

\section{Bosonization}\label{sec:3}

\vspace*{-.3cm}

\subsection{Bosonized Hamiltonian}

\vspace*{-.3cm}

The dominant low-energy behavior of the model defined by
Eqs.\ (\ref{eq:hamilt}) can be studied by using a harmonic fluid
representation, where the single-particle dispersion relations are
linearized near the ``Fermi'' points.  In the problem of BEC-BCS
crossover in one dimension, a two-channel model of two-component
fermions that dimerize into bosonic molecules has been previously
analyzed by means of the bosonization method in
Refs.\ \cite{Sheehy2005} and \cite{Orignac:2006cb}.
Because of the different statistics of particles, the bosonization
analysis of the present model will reveal different phases.

In terms of the bosonic phase fields $\phi_s(x)$, 
the density
operators can be expressed as
\cite{giambook,Cazalilla:2004dd,Cazalilla_review}
\begin{equation}
 \rho_s (x) = \rho^0_s  - 
\frac{1}{\pi} \frac{d\phi_s(x)}{dx}
+ \rho^0_s
\sum_{m\neq 0}
e^{2im[\pi\rho^0_s x-\phi_s(x)]}
,
\label{density}
\end{equation}
where $\rho^0_s$ is the equilibrium density
and the summation is over nonzero integer $m$.
The field operators for the respective particles are represented as
\cite{giambook,Cazalilla:2004dd,Cazalilla_review}
\begin{subequations}
\begin{eqnarray}
 \Psi_b  (x) 
\!\! &=& \!\!
\frac{1}{\sqrt{2\pi\alpha}}
\sum\limits_{n\in\mathbb{Z}}
e^{in[ 2\pi \rho^0_b x - 2\phi_b (x) ]
 + i\theta_b (x)},\qquad
\label{eq:psi_b}
\\
 \Psi_f^{L/R}  (x) 
\!\! &=& \!\!
\frac{\xi_f}{\sqrt{2\pi\alpha}}
e^{\mp i[ {\pi \rho^0_f x - {\phi _f }(x)} ] + i{\theta_f }(x)},
\\
 \Psi_\psi^{L/R}  (x) 
\!\! &=& \!\!
\frac{\xi_\psi}{\sqrt{2\pi\alpha}}
e^{\mp i[ {\pi \rho^0_\psi x - {\phi _\psi}(x)} ] + i{\theta_\psi }(x)},
\qquad
\end{eqnarray}%
\label{eq:psi}%
\end{subequations}
where $\alpha$ is a short-distance cutoff.
The field operators
$\Psi_s^{L}$ and $\Psi_s^{R}$ ($s=f,\psi$) represent the
left-moving and right-moving
chiral branches of fermionic particles, respectively.
The Klein factors 
$\xi_f$ and $\xi_\psi$, satisfying 
  $\{\xi_s,\xi_{s'}\}=2\delta_{s,s'}$ and $\xi_s^\dagger =\xi_s$,
are introduced in order to retain the anticommutation relation 
between $f$ and $\psi$ particles.
 $\theta_s(x)$ are dual fields to $\phi_s(x)$ and obey
$\left[\phi_s (x),\theta_{s'}(x')\right]
 = i\pi\delta_{s,s'}\Theta(-x + x')$,
where $\Theta(x)$ is the Heaviside step function,
i.e., $\Theta(x)=1$ for $x>0$, $\Theta(0)=\frac12$,
and $\Theta(x)=0$ for $x<0$.
By introducing the conjugate
momenta $\Pi_s(x) = (1/\pi)\partial_x\theta_s (x)$,
a generic TLL Hamiltonian for each component is expressed as
\begin{equation}
H_{s} 
= \frac{u_s}{2\pi}
\int {dx} \left\{
K_{s}[\pi \Pi_{s} (x)]^2 
+ \frac{1}{K_{s}}[\partial_x \phi_{s} (x)]^2
\right\}.
\label{eq:H0}
\end{equation}
The parameters $u_s$ and $K_s$ are velocities and TLL parameters,
respectively,
which depend on the precise forms of microscopic intraspecies
interactions.
We will consider the general case where $0<K_{s=b,f,\psi}< \infty$.
The noninteracting limits $V_{bb}\to 0$ and $V_{ff},\,\, V_{\psi \psi} \to 0$ 
correspond to $K_b=\infty$ and $K_{s=f,\psi}=1$, respectively.
By tuning $g_{b} \to \infty$, the system enters the TG regime at
$K_b \gtrsim 1$ \cite{giambook,Cazalilla_review}.
For specific realizations of optical lattice systems,
the commensurability of the Bose-Hubbard interaction allows the
possibility of tuning into the regime $K_b < 1$, when $V_{bb}$ is long ranged
\cite{giambook,Cazalilla_review}.

After substituting the bosonized form of $\Psi_s(x)$ defined in
Eqs.\ (\ref{eq:psi}) into Eq.\ (\ref{eq:hamilt-g3})
and keeping only the $n=0$ term for $\Psi_b$, we obtain
\begin{eqnarray}
H_{3p} \!\! &=& \!\!
-i \tilde g_{3p}
\int dx 
\cos[\theta_b(x) + \theta_f(x) - \theta_\psi(x)]
\nonumber \\ && {} \qquad\qquad \times
 \sin[\phi_f(x) - \phi_\psi (x) - \delta k_F x ], \quad
\label{eq:Hg3p}
\end{eqnarray}
where $\delta k_F \equiv k_F^f-k_F^\psi$ and $\tilde
g_{3p}=4g_{3p}(2\pi\alpha)^{-3/2}$.
In deriving Eq.\ (\ref{eq:Hg3p}) 
we have discarded terms like
$\tilde g_{3p} \sin (\theta_b + \theta_f - \theta_\psi) \cos( 2 k_F x
- \phi_f - \phi _\psi )$,
which are strongly irrelevant in the RG sense because they have
spatial oscillations with the wave number
$2k_F\equiv k_F^f + k_F^\psi =\pi\mathcal{N}_F/L$.
Furthermore, we have replaced $\xi_f\xi_\psi$
with $+i$, because the product of the two Klein factors is a constant
of motion
[the identity $(\xi_f\xi_\psi)^2=-1$ implies either $\xi_f\xi_\psi=+i$
or $-i$, and we have chosen the former].
We will use the same sign convention when we derive bosonized form of
order parameters.

In the incommensurate case ($\delta k_F \neq 0$), the $g_{3p}$
interaction [Eq.\ (\ref{eq:Hg3p})] is irrelevant 
in the RG sense.
The analysis of the previous section is then applicable, and the phase
diagram therefore can be determined as in the previous section,
with various depletion transitions occurring between binary and
ternary mixture phases.
On the other hand, Eq.\ (\ref{eq:Hg3p}) has a dramatically
different effect for the commensurate case ($\delta k_F = 0$),
which is satisfied along the dashed line in Fig.~\ref{fig:pd-limit}.
In this case, sinusoidal potentials can lock a particular phase variable
($\theta_s$ or $\phi_s$), and a competition of various
orders due to the phase locking must be studied by performing a RG
analysis.

\vspace*{-.3cm}

\subsection{Order parameters}
\label{section:OP}

\vspace*{-.3cm}

In the context of quantum mixtures, composite ``pairing''  correlations
have been previously introduced in the literature and will be extended
here to a more comprehensive list of possible order parameters.
First, the conventional $2k_F$ density-wave (DW) order parameters
are given by
\begin{subequations}
\begin{eqnarray}
\mathcal{O}^\mathrm{DW}_b (x) 
\!\! &=& \!\!
\Psi_b^\dagger(x) \Psi_b(x) 
\simeq 
e^{i(2k_F^bx-2\phi_b)},
\quad
\\
\mathcal{O}^\mathrm{DW}_f (x) 
\!\! &=& \!\!
\Psi_f^{L\dagger}(x) \Psi_f^R(x)
\simeq 
e^{i(2k_F^fx-2\phi_f)},
\\
\mathcal{O}^\mathrm{DW}_\psi (x) 
\!\! &=& \!\!
\Psi_\psi^{L\dagger}(x) \Psi_\psi^R(x)
\simeq 
e^{i(2k_F^\psi x-2\phi_\psi)}.
\end{eqnarray}
Here (and below) we have dropped unimportant numerical prefactors.
In analogy with order parameters in the spinless two-coupled chain system 
[see Eq.\ (\ref{eq:2chain-O-CDW})],
we introduce the out-of-phase DW state of $f$ and $\psi$ particles,
\begin{eqnarray}
\mathcal{O}^{\mathrm{DW}}_{f\psi}(x)
\!\!&=&\!\! 
\Psi_f^{L\dagger}(x) \Psi_f^R(x) - \Psi_\psi^{L\dagger}(x) \Psi_\psi^R(x)
\nonumber \\
&\simeq&\!\!
e^{i2k_Fx -i(\phi_f+\phi_\psi)} 
 \sin (\phi_f-\phi_\psi-\delta k_F x).
\nonumber \\
\label{eq:O_DW_fpsi}
\end{eqnarray}
Next, the order parameters for the superfluidity (SF) of bosons, 
$p$-wave-paired fermions, and $p$-wave-paired molecules 
are given by
\begin{eqnarray}
\mathcal{O}^\mathrm{SF}_b (x)
\!\! &=& \!\!
\Psi_b (x) 
\simeq e^{i\theta_b},
\\
\mathcal O^\mathrm{SF}_{ff} (x) 
\!\! &=& \!\!
\Psi_f^L(x)\Psi_f^R(x) \simeq e^{i2\theta_f},
\\
\mathcal O^\mathrm{SF}_{\psi\psi} (x) 
\!\! &=& \!\!
\Psi_\psi^L(x)\Psi_\psi^R(x) \simeq e^{i2\theta_\psi}.
\end{eqnarray}
We will also consider the $p$-wave-paired SF state 
composed of $f$ and $\psi$ particles,
\begin{eqnarray}
\mathcal{O}^{\mathrm{SF}}_{f\psi}(x)
\!\! &=& \!\!
\Psi_f^L(x) \Psi_\psi^R(x) - \Psi_f^R(x) \Psi_\psi^L(x)
\nonumber \\
&\simeq& 
 e^{i(\theta_f+\theta_\psi)} 
 \sin (\phi_f-\phi_\psi-\delta k_F x), \qquad
\label{eq:fpsi}
\end{eqnarray}
which is odd under the parity transformation, $L\leftrightarrow R$, and therefore can be classified as $p$-wave pairing.
Moreover, this order parameter can be identified with the
interchain pairing SC$d$ state in the two-coupled spinless chain
problem [see Eq.\ (\ref{eq:2chain-O-SCd})], where $f$ and $\psi$ can be replaced by the two-chain indices.

Earlier work in Ref.\ \cite{Marchetti2009} investigated the phase
diagram of interacting ``$b$+$f$'' binary mixtures near the commensurate
point $\bar \rho_b=\bar \rho_f$, where the composite $p$-wave
superfluidity $\sim \Psi_b^2 \Psi_f^L \Psi_f^R$ was shown to have
dominant QLRO correlations. 
In the ternary system studied here,
similar orders can persist,
\begin{eqnarray}
\mathcal{O}^{\mathrm{SF}}_{bff+b^\dagger\psi\psi}(x)
\!\! &=& \!\!
\Psi_b \Psi_f^L \Psi _f^R 
+
\Psi_b^\dag \Psi_\psi^{L} \Psi_\psi^{R}
\nonumber \\
&\simeq& 
e^{i(\theta_f + \theta_\psi)}
\cos (\theta_b + \theta_f - \theta_\psi) ,
\qquad
\label{eq:Obff}
\end{eqnarray}
which describes 
the $p$-wave pairing of two fermionic ($f$ or $\psi$) particles 
combined with a single $b$ atom.
Note that the two composite operators in Eq.\ (\ref{eq:Obff}),
$\Psi_b \Psi_f^L \Psi _f^R$ and
$\Psi_b^\dag \Psi_\psi^{L} \Psi_\psi^{R}$,
annihilate equal numbers of fermionic and bosonic
atoms (including the ones forming a molecule),
as seen from the commutation relations
$[\mathcal{N}_B,\mathcal{O}^{\mathrm{SF}}_{bff+b^\dagger\psi\psi}]=
-\mathcal{O}^{\mathrm{SF}}_{bff+b^\dagger\psi\psi}$ and
$[\mathcal{N}_F,\mathcal{O}^{\mathrm{SF}}_{bff+b^\dagger\psi\psi}]=
-2\mathcal{O}^{\mathrm{SF}}_{bff+b^\dagger\psi\psi}$.
This order parameter corresponds to the intrachain SC$s$ pairing
in the two-coupled chain problem
[see Eq.\ (\ref{eq:2chain-O-SCs})].

\begin{figure*}[t]
\centerline{
\includegraphics[width=7.5cm]{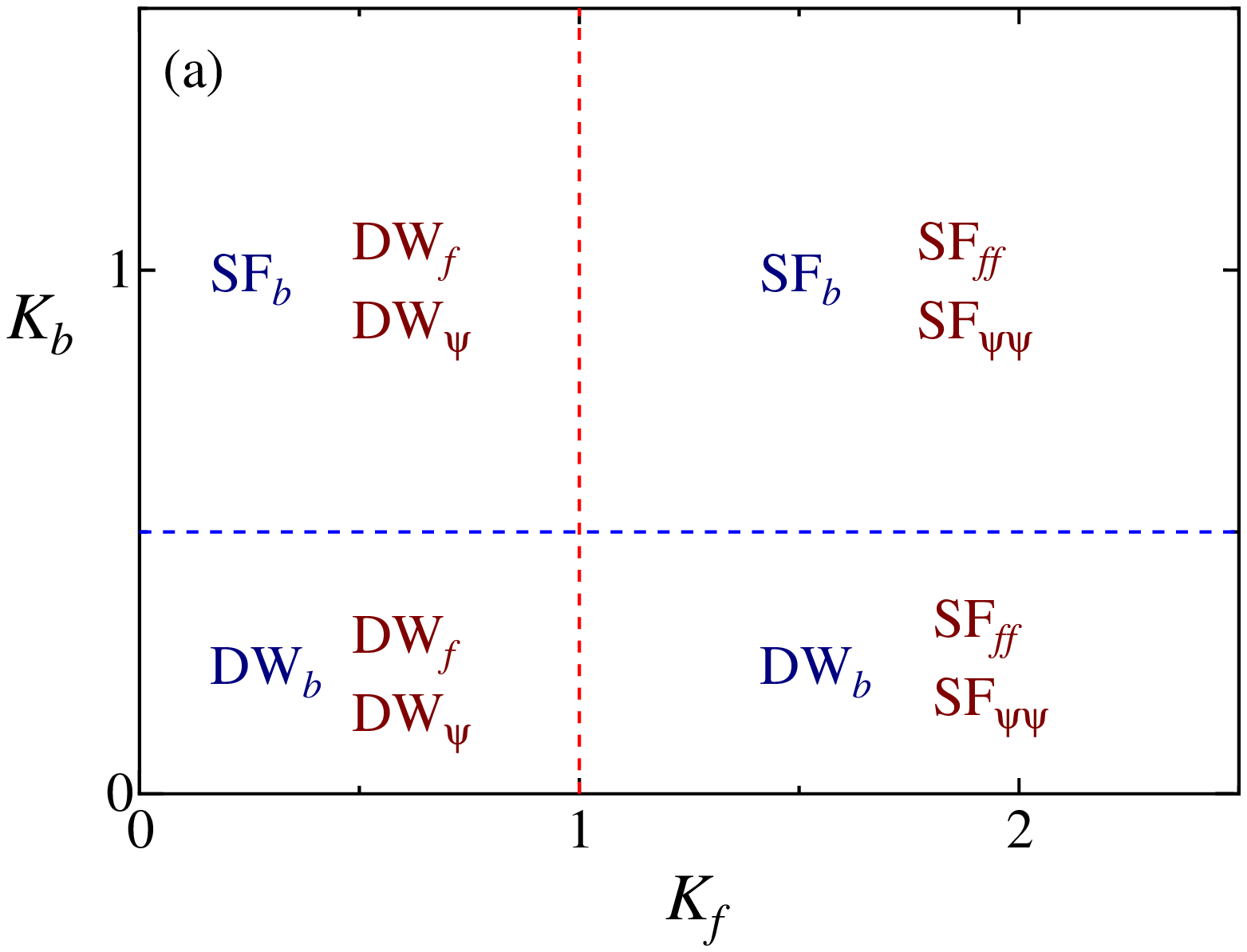}
\hspace*{.5cm}
\includegraphics[width=7.5cm]{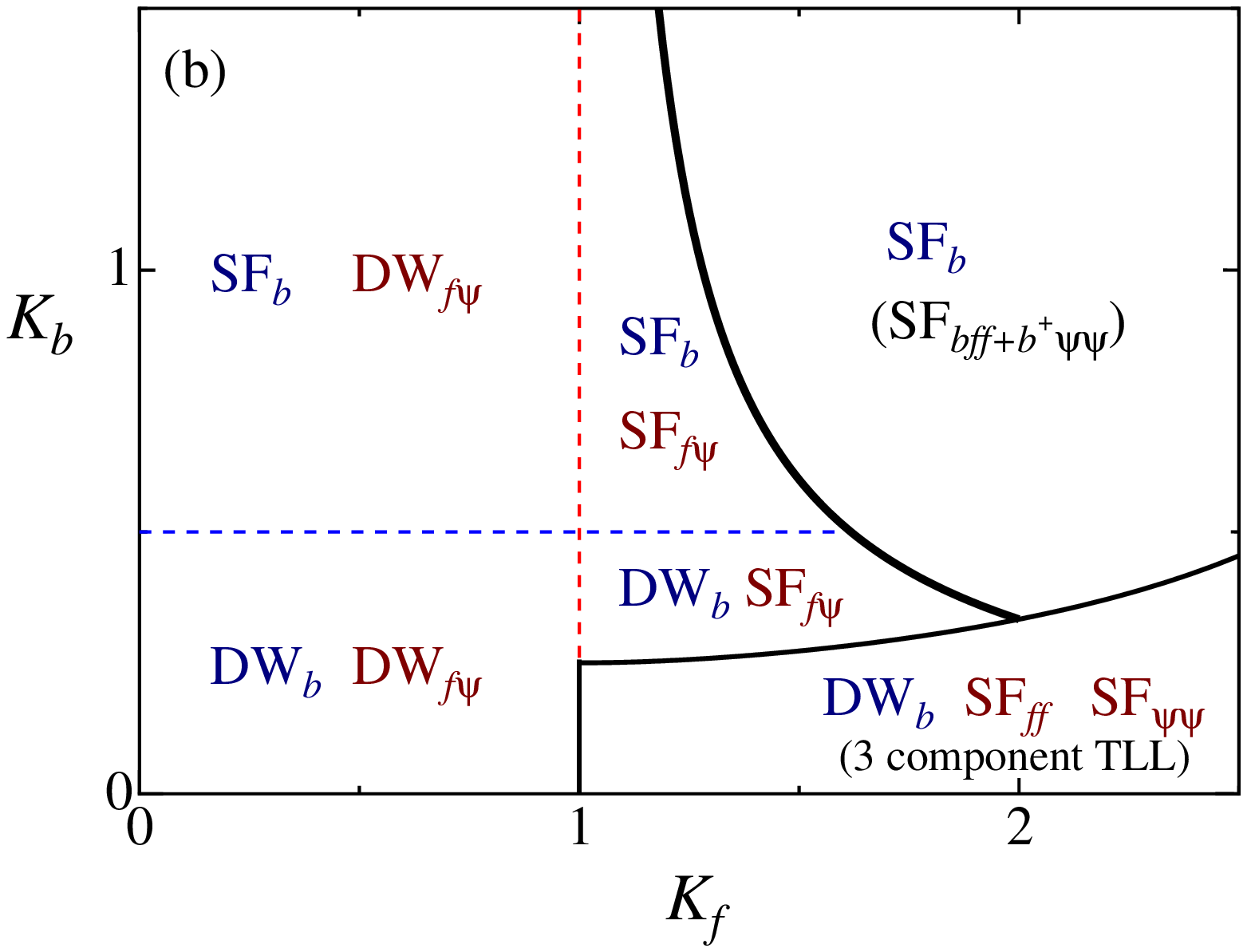}
}
\caption{
(Color online)
Phase diagram of Hamiltonian (\ref{eq:fullham})
for the incommensurate case $\rho_f\neq \rho_\psi$ (a) and the
commensurate case $\rho_f=\rho_\psi$ (b).
For simplicity we set $u_b=u_f=u_\psi$ and $K_\psi=K_f$. 
The regions labeled by DW and SF
represent phases with the dominant density-wave and superfluid correlations,
respectively.
The dominant correlation crosses over from DW to SF or vice versa
across the dashed lines.
(b)
In the phase denoted by ``(3 component TLL),''
all the couplings $G_{3p}$, $G_{\phi}$, and
$G_{\theta}$ are irrelevant in the RG sense.
The boundary between the phases of relevant $G_\phi$ and that of
relevant $G_\theta$ is shown by the thick solid line
at which the system undergoes a quantum phase transition. 
On the left-hand (right-hand) side of the thick solid line, 
the coupling $G_\phi$ ($G_\theta$) becomes relevant.
}
\label{fig:pd}
\end{figure*}

In addition, we consider other composite order parameters
defined by
\begin{eqnarray}
\mathcal{O}^\mathrm{ph1}_{b^\dagger f^\dagger \psi} (x)
\!\! &=& \!\! 
\Psi_b^\dagger \Psi_f^{L\dagger} \Psi_\psi^L 
-\Psi_b \Psi_\psi^{R\dagger} \Psi_f^R 
\nonumber \\
&\simeq& \!\!
 e^{i\delta k_F x-i(\phi_f-\phi_\psi) }
 \sin (\theta_b+\theta_f-\theta_\psi)
\nonumber \\ && {} \!\!
+
 e^{i(-2k_F^b+\delta k_F) x+i( 2\phi_b-\phi_f+\phi_\psi) }
\nonumber \\&& {}  \quad \times
 \cos(\theta_b+\theta_f-\theta_\psi),
\label{eq:Obfp1}
\\ 
\mathcal{O}^\mathrm{ph2}_{b^\dagger f^\dagger \psi} (x)
\!\! &=& \!\!
\Psi_b^\dag \Psi_f^{L\dagger} \Psi_\psi^R
-\Psi_b \Psi_\psi^{L\dagger} \Psi_f^R 
\nonumber \\
&\simeq& \!\!
 e^{i2k_F x -i(\phi_f+\phi_\psi) }
\cos (\theta_b+\theta_f-\theta_\psi),
\qquad
\label{eq:Obfp2}
\end{eqnarray}%
\label{eq:OP}%
\end{subequations}
which represent the particle-hole combinations of $f$ and $\psi$ fermions.
These operators are composed of the products of three field operators,
 $\Psi_b^\dagger\Psi_f^\dagger\Psi_\psi$ and
 $\Psi_b\Psi_\psi^\dagger\Psi_f$,
which are similar in form to the $g_{3p}$ term 
of Eq.\ (\ref{eq:hamilt-g3}) but asymmetrical in the $L,R$ branches.
The second bosonized contribution in Eq.\ (\ref{eq:Obfp1}),
coming from the $n= - 1$ contribution in Eq.\ (\ref{eq:psi_b}),
can become a dominant order parameter for some parameter regime,
as will be shown later.
We also note that the order parameter in Eq.\ (\ref{eq:Obfp2}) corresponds
to the ``orbital antiferromagnetic state'' in the two-coupled chain
problem, in which circulating currents flow between the two chains,
if the $f$ and $\psi$ indices are regarded as chain indices
[see Eq.\ (\ref{eq:2chain-O-OAF})].

\vspace*{-.3cm}

\subsection{Ground states in the incommensurate case}

\vspace*{-.3cm}

When $\delta k_F\ne0$,
the $g_{3p}$ interaction of Eq.\ (\ref{eq:Hg3p})
oscillates in space and does not affect the low-energy spectrum. 
Thus we can set $g_{3p}=0$ in the low-energy limit, and 
the system is described as a three-component TLL,
in which the $b$, $f$, and $\psi$ particles are decoupled
and the correlation functions exhibit algebraic decay.
For example, the correlators for the $b$ particles are given by
\begin{subequations}
\begin{eqnarray}
\langle \mathcal{O}^{\mathrm{SF}}_b(x) \,
    \mathcal{O}^{\mathrm{SF}\dag}_b(0) \rangle_{0} 
&\sim&
 x^{-1/(2K_b)},
\\
\langle \mathcal{O}^\mathrm{DW}_b(x) \,
  \mathcal{O}^\mathrm{DW^\dag}_b(0) \rangle_{0}
&\sim&
 x^{-2K_b} e^{i2k_F^b x}. \quad
\end{eqnarray}%
\label{eq:cor0b}%
\end{subequations}
We find that the superfluidity correlation dominates over the
density-wave correlation when $K_b>1/2$.
Similarly, the correlation functions for the $p$-wave
superfluidity and the density-wave of the $f$ and $\psi$ particles
exhibit algebraic decay,
\begin{subequations}
\begin{eqnarray}
\langle \mathcal{O}^\mathrm{SF}_{ss}(x) \,
  \mathcal{O}^{\mathrm{SF}\dag}_{ss}(0) \rangle_0 
&\sim&
 x^{-2/K_s},
\\
\langle \mathcal{O}_s^{\mathrm{DW}}(x) \,
  \mathcal{O}_s^{\mathrm{DW}\dagger}(0) \rangle_0
&\sim&
 x^{-2K_s} e^{i2k_F^s x}, \quad
\end{eqnarray}
\label{eq:cor0f}%
\end{subequations}
where $s=f,\psi$.
The dominant correlation for fermions changes between
the superfluidity and density-wave orders at $K_s=1$.
In Fig.\ \ref{fig:pd}(a) we show the phase diagram in the parameter
space of $K_s$ ($s=b, f, \psi$), which is obtained
by identifying the dominant QLRO among those in Eqs.\ (\ref{eq:cor0b})
and (\ref{eq:cor0f}).

\vspace*{-.3cm}

\section{Renormalization in the commensurate case}\label{sec:RG}

\vspace*{-.3cm}

When $\delta k_F\simeq 0$, the effects of the sinusoidal potential
(\ref{eq:Hg3p}) can be analyzed using RG techniques \cite{giambook}. 
Apparently, the form of Eq.\ (\ref{eq:Hg3p}) 
contains dual fields which do not commute 
 $[\theta_b+\theta_f-\theta_\psi,\phi_f-\phi_\psi]\neq 0$.
This type of interaction has been analyzed 
in the context of two TLL chains coupled 
by one-particle interchain hopping \cite{gogolinbook,nersesyanPLA1993}, 
where it has been confirmed that higher-order corrections
are crucial to determine the low-energy spectrum
of the two TLL chains \cite{yakovenkoJETP1992}.
We thus can expect that interactions generated by
RG transformation should similarly be taken into account in our model.

In order to properly derive the RG equations and to
determine the ground-state phase diagram
for the Hamiltonian including the potential as Eq.\ (\ref{eq:Hg3p}), 
we have to pay special attention to the commutative properties of the 
phase fields, besides the Klein factors.
In Appendix \ref{sec:app-A}, we analyze the two-coupled chain
system on the basis of the present bosonization scheme and 
verify that the correct results \cite{Orignac:1997vq} can be derived.
In Ref.\ \cite{Orignac:1997vq}, 
the interchain hopping term was treated nonperturbatively
and the phase diagram was determined.
In Appendix \ref{sec:app-B}, we analyze the present model (\ref{eq:fullham})
using the method of Ref.\ \cite{Orignac:1997vq}
and observe that the consistent results can be obtained.

In this section 
we set $u_b=u_f=u_\psi(\equiv u)$ for simplicity.
The Euclidean action of the system is given by
$S= S_0+S_{I,0}+S_{I,1}+S_{I,2}+S_{I,3}$
with
\begin{subequations}
\begin{eqnarray}
S_0 &=& \sum_s \frac{1}{2\pi K_s}\int d^2r
\left(\nabla \phi_s\right)^2 ,
\\
S_{I,0}
\!\!&=&\!\! 
 \sum_{s\neq s'}\frac{G_{ss'}}{2\pi} 
\int d^2 r \,  (\nabla \phi_s) (\nabla \phi_{s'}),
\\
S_{I,1}
\!\!&=&\!\! 
\frac{G_{3p}}{i\pi}
\int \frac{d^2 r}{\alpha^2}
\cos(\theta_b + \theta_f - \theta_\psi)
\sin( \phi_f - \phi_\psi) ,
\nonumber \\
\\
S_{I,2}
\!\!&=&\!\! 
\frac{G_{\phi}}{\pi}
\int \frac{d^2 r}{\alpha^2}
\cos(2 \phi_f - 2\phi_\psi) ,
\\
S_{I,3}
\!\!&=&\!\! 
\frac{G_{\theta}}{\pi}
\int \frac{d^2 r}{\alpha^2}
\cos(2\theta_b + 2\theta_f - 2\theta_\psi),
\end{eqnarray}%
\label{eq:action}%
\end{subequations}
where $\bm r=(x,u\tau)$,
$\nabla=(\partial_x , u^{-1} \partial_\tau)$,
$d^2r = udx d\tau$, and
 $G_{3p}=\pi\alpha^2 \tilde g_{3p}/u$.
Although the extra terms $G_{\phi}$, $G_\theta$, $G_{bf}$, $G_{b\psi}$,
and $G_{f\psi}$, are absent in the original Hamiltonian, they are generated
through the RG process \cite{nersesyanPLA1993}.

In this paper, we adopt the momentum-space RG method
\cite{Kogut:1979wg} by introducing the momentum space cutoff $\Lambda$.
The RG equations can be obtained by integrating out the high-momentum
components $\Lambda' < |\bm{k}| < \Lambda$, where 
$\Lambda'=\Lambda(1-dl)$ is the reduced cutoff
($dl=-d\Lambda/\Lambda$) and $\bm k=(k,\omega/u)$ with the
frequency $\omega$.
Accordingly, the phase fields $\phi_s(\bm r)$ are split into two 
components $\phi_s(\bm r)=\phi_s'(\bm r)+h_s(\bm r)$ 
 \cite{Kogut:1979wg}, where 
$\phi'_s(\bm r)$ is the field having components in lower momentum
$0< |\bm k| < \Lambda'$ and $h_s(x)$ has higher momentum components
$\Lambda'<|\bm k| <\Lambda$.
The free propagators for these fields are given by
\begin{eqnarray}
\langle \phi_s'(\bm r)\phi_s'(\bm 0) \rangle
\!\! &=& \!\!
\frac{K_s}{2} \bar g(r)
=
 \frac{K_s}{2} \int_0^\infty \frac{dk}{k} 
J_0(kr) f(k/\Lambda') , 
\nonumber \\
\label{eq:phipcorre}
\\
\langle h_s(\bm r)h_s(\bm 0) \rangle
\!\! &=& \!\!
\frac{K_s}{2} \delta g(r) 
\nonumber \\
&=& \!\! \frac{K_s}{2} \int_0^\infty \frac{dk}{k} 
J_0(kr)
[f(k/\Lambda)-f(k/\Lambda')],
\nonumber \\
\label{eq:hcorre}
\end{eqnarray}
where $r=|\bm r|$ and
$J_0(z)$ is the Bessel function of the first kind.
With the smooth cutoff function $f(p)=c^2/(p^2+c^2)$ \cite{Ohta:1979tg},
we have the correlation functions 
$\langle [\phi'_s(\bm r)-\phi'_s(\bm 0)]^2\rangle
   = K_s \ln(e^\gamma c\Lambda' |\bm r|/2)$ 
and 
$\delta g(r) = c\Lambda r K_1(c\Lambda r) dl$
for $c\Lambda r\gg 1$, where 
$K_1(z)$ is the modified Bessel function.
The constant $c$ is taken as
$c=2e^{-\gamma}/(\Lambda\alpha)$ in order to reproduce
the asymptotic form
$\langle [\phi_s(\bm r)-\phi_s(\bm 0)]^2\rangle
  = K_s \ln(|\bm r|/\alpha)$ for $|\bm r|\to\infty$.
The derivation of one-loop RG equations proceeds similarly to
the case of the two-coupled chain system explained in Appendix \ref{sec:app-A}.
By exploiting the commutation relation
$[\phi_s(x),\theta_{s'}(x')]=i\pi \delta_{s,s'}\Theta(-x+x')$
and the normal ordering procedure for the operator-product expansion
\cite{giambook,Nozieres:1987kg},
we eventually obtain the following one-loop RG equations:
\begin{subequations}
\begin{eqnarray}
\frac{d G_{3p}}{dl} \!\!&=&\!\!
\Bigl(2-\frac{1}{4K_b}-\frac{1}{4K_f}-\frac{1}{4K_\psi}
  -\frac{K_f}{4}-\frac{K_\psi}{4}
- \frac{1}{2} G_{bf}
\nonumber \\ && {} \!\!\!\!
+ \frac{1}{2} G_{b\psi}
+ \frac{1}{2} G_{f\psi} 
- \frac{1}{2} G_{f\psi} K_fK_\psi 
\Bigr) \!
G_{3p},
\label{eq:RG-1}
\\
\frac{d G_\phi}{dl} \!\!&=&\!\! 
\bigl(2-K_f-K_\psi -2 G_{f\psi}K_fK_\psi \bigr)  G_\phi
\nonumber \\ && {} \hspace*{-.5cm}
 + \frac{1}{4} G^2_{3p}  \,
A_1\biglb((K_b^{-1}+K_f^{-1}+K_\psi^{-1}-K_f-K_\psi)/4\bigrb),
\nonumber \\ && {}
\label{eq:RG-2}
\\
\frac{d G_\theta}{dl} \!\!&=&\!\!
\Bigl(2-\frac{1}{K_b}-\frac{1}{K_f}-\frac{1}{K_\psi}
- 2G_{bf}
\nonumber \\ && {} 
 +2G_{b\psi} +2G_{f\psi}\Bigr) \! G_\theta
\nonumber \\ && {} \hspace*{-.5cm}
 -  \frac{1}{4} G^2_{3p} \,
A_1\biglb((K_f+K_\psi-K_b^{-1}-K_f^{-1}-K_\psi^{-1})/4\bigrb),
\nonumber \\ && {} \!\!\!
\label{eq:RG-3}
\\
\frac{d K_{b}}{dl} \!\!&=&\!\!{} 
+G^2_{\theta} A_2(K_b^{-1}+K_f^{-1}+K_\psi^{-1}),
\label{eq:RG-4}
\\
\frac{d K_{f}}{dl} \!\!&=&\!\!{} 
-  G_\phi^2 \, K_f^2 \, A_2(K_f+K_\psi)
\nonumber \\ && {} \!\!\!
+  G^2_{\theta}A_2(K_b^{-1}+K_f^{-1}+K_\psi^{-1}),
\label{eq:RG-5}
\\
\frac{d K_{\psi}}{dl} 
\!\!&=&\!\!{} 
- G_\phi^2 \, K_\psi^2 \, A_2(K_f+K_\psi)
\nonumber \\ && {} \!\!\!
+ G^2_{\theta} \,  A_2(K_b^{-1}+K_f^{-1}+K_\psi^{-1}),
\label{eq:RG-6}
\\
\frac{d G_{bf}}{dl} \!\!&=&\!\!{} 
+ \frac{G^2_{\theta}}{K_b K_f} \, A_2(K_b^{-1}+K_f^{-1}+K_\psi^{-1}),
\\
\frac{d G_{b\psi}}{dl} \!\!&=&\!\!{} 
+ \frac{G^2_{\theta}}{K_b K_\psi} \,
 A_2(K_b^{-1}+K_f^{-1}+K_\psi^{-1}),
\\
\frac{d G_{f\psi}}{dl} \!\!&=&\!\!{} 
- G_\phi^2 \, A_2(K_f+K_\psi)
\nonumber \\ && {} \!\!\!
+ \frac{G^2_{\theta}}{K_f K_\psi}\,
  A_2(K_b^{-1}+K_f^{-1}+K_\psi^{-1}) ,
\end{eqnarray}%
\label{eq:RG}%
\end{subequations}
where we have defined
\begin{subequations}
\begin{eqnarray}
A_1(\beta) dl
\!\! &\equiv& \!\!
2\beta
\int_0^\infty \frac{dr}{\alpha}\frac{r}{\alpha}
 \delta g(r) 
e^{-2\beta[\bar g(0) - \bar g(r)]},
\\
A_2(\beta) dl
\!\! &\equiv& \!\!
2\beta
\int_0^\infty \frac{dr}{\alpha}\frac{r^3}{\alpha^3}
 \delta g(r) e^{-2\beta[\bar g(0) - \bar g(r)]} .
\qquad
\end{eqnarray}%
\label{eq:A}%
\end{subequations}
The exponential factors in the rhs of Eqs.\ (\ref{eq:A}) appear 
as a result of normal ordering
in operator-product expansions \cite{giambook,Nozieres:1987kg};
for example,
\begin{eqnarray}
&& \hspace*{-.6cm}
 \cos \left[
            p\phi'_{s}(\bm r_1) + q\phi'_{s} (\bm r_2)
       \right]
\nonumber \\
&=& \!\! 
 : \! \cos\!\left[
            p\phi'_{s}(\bm r_1) \! + \! q\phi'_{s} (\bm r_2)
       \right] \! : \,
e^{-  \frac{1}{2}(p^2+q^2)  \langle {\phi'_{s}}^2\rangle 
            -  pq \langle \phi'_{s}(\bm 1)\phi'_{s}(\bm 2)\rangle}
\nonumber \\
&\approx& \!\! 
 : \! \cos\!\left[
            (p \! + \! q)\phi'_{s}(\bm R)
       \right] \! : 
e^{-  \frac{1}{2}(p^2+q^2)  \langle {\phi'_{s}}^2\rangle 
            -  pq \langle \phi'_{s}(\bm 1)\phi'_{s}(\bm 2)\rangle}
\nonumber \\
&=& \!\! 
 \cos\!\left[
            ( \! p + \! q)\phi'_{s}(\bm R)
       \right]  \,
e^{\frac{1}{2}[(p+q)^2 - (p^2+q^2)] \langle {\phi'_{s}}^2\rangle
            -  pq \langle \phi'_{s}(\bm 1)\phi'_{s}(\bm 2)\rangle}
\nonumber \\
&=& \!\! 
 \cos\!\left[ (p+q)\phi'_{s}(\bm R) \right] \,
e^{\frac{1}{2} pq K [\bar g(0) - \bar g (r_{12})]},
\label{eq:ope}
\end{eqnarray}
where $\bm R=(\bm r_1 + \bm r_2)/2$ and
$r_{12} = |\bm r_1 - \bm r_2|$.
We have used the short-hand notations $\bm 1=\bm r_1$ and
$\bm 2=\bm r_2$.
We note that $A_1(\beta)\approx e^{2\gamma} \beta$ for
small $\beta$, and $A_1(1)=A_2(2)=1$, where 
$\gamma$ is the Euler-Mascheroni constant. 
One can neglect the velocity renormalization up to
one-loop order.
The initial values of the RG equations are given by
$G_{3p}(0)=G_{3p}$, $K_s(0)=K_s$, and 
$G_{ss'}(0)=G_{\phi}(0)=G_\theta(0)=0$.

\begin{figure}[t]
\centerline{\includegraphics[width=7.5cm]{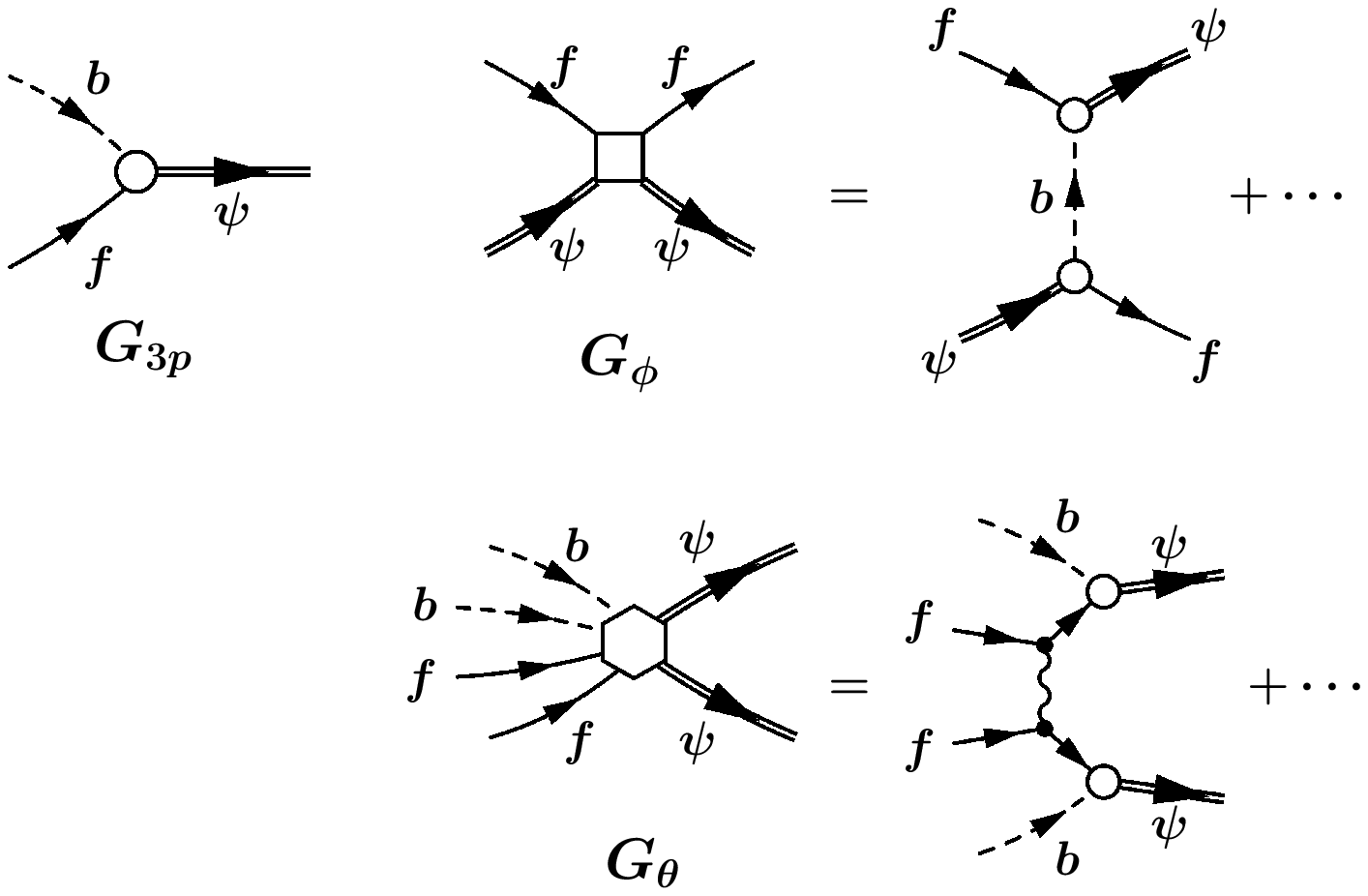}}
\caption{
Diagrammatic representation of the 
$G_{3p}$, $G_\phi$, and $G_\theta$ terms, and low-order contributions
to $G_\phi$ and $G_\theta$.
The dashed, sold, and double lines represent the boson, fermion, and
molecule propagators, respectively.
The wavy line represents the intraspecies density-density interaction.
}
\label{fig:vertex}
\end{figure}

Diagrammatic representations for the $G_{3p}$, $G_\phi$, and $G_\theta$
terms are shown in Fig.\ \ref{fig:vertex}.
The $G_\phi$ coupling is a four-point vertex representing 
interactions between 
$f$ and $\psi$ particles, while the $G_\theta$ coupling is a six-point
vertex for a two-molecule conversion from two $b$ and two $f$ particles.
Low-order contributions to $G_\phi$ and $G_\theta$ are also shown in 
Fig.\ \ref{fig:vertex}.
The lowest-order contribution to the $G_\phi$ coupling
comes from the effective interaction mediated by $b$ atoms.
Pairing between fermions ($f$) and molecules ($\psi$) induced by 
such boson- ($b$) mediated interaction has been suggested in
Ref.\ \cite{Zhang:2005dr}.
We will contrast this paper with our work in more detail later.

Since $[\phi_f-\phi_\psi, \theta_b+\theta_f-\theta_\psi]\neq 0$,
the phase variables $\phi_f-\phi_\psi$ and
$\theta_b+\theta_f-\theta_\psi$
cannot be locked simultaneously.
This means that there should be two distinct phases
separated by a quantum phase transition,
a phase where $\phi_f-\phi_\psi$ is locked by
the $G_\phi$ term and a phase where $\theta_b+\theta_f-\theta_\psi$
is locked by the $G_\theta$ term,
in addition to a three-component TLL phase
where none of the phase fields are locked.
Here we obtain the phase diagram by comparing the scaling dimensions,
which we denote by dim[\ ], of the operators for the couplings
$G_{3p}$, $G_\phi$, and $G_\theta$.
We ignore renormalization of $K_b$, $K_f$, and $K_\psi$ for weak $g_{3p}$,
because the
right-hand side of Eqs.\ (\ref{eq:RG-4})$-$(\ref{eq:RG-6}) 
are of order $g_{3p}^4$.
The scaling dimensions of the sinusoidal potential operators are
found from Eqs.\ (\ref{eq:RG-1})$-$(\ref{eq:RG-3}) as
\begin{eqnarray}
\mathrm{dim}[G_{3p}]\!\!&=&\!\!
\frac{1}{4}\!\left(\frac{1}{K_b}+\frac{1}{K_f}+\frac{1}{K_\psi}
                 +K_f+K_\psi\right),
\nonumber \\
\mathrm{dim}[G_{\phi}]\!\!&=&\!\! K_f+K_\psi,
\\
\mathrm{dim}[G_{\theta}]\!\!&=&\!\!
\frac{1}{K_b}+\frac{1}{K_f}+\frac{1}{K_\psi}.
\nonumber 
\end{eqnarray}
In the case when three inequalities,
$\mathrm{dim}[G_{3p}]>2$, $\mathrm{dim}[G_\phi]>2$, and
$\mathrm{dim}[G_\theta]>2$, are simultaneously satisfied,
all the locking potential operators are irrelevant, and
consequently, we have a three-component TLL phase.
This is the case for large $K_f$ and $K_\psi$ and small $K_b$.
Otherwise,
either the coupling $G_\phi$ or $G_\theta$ becomes relevant
and flows to strong coupling at low energy.

We observe from Eqs.\ (\ref{eq:RG-2}) and (\ref{eq:RG-3}) that the condition
\begin{equation}
K_f+K_\psi=\frac{1}{K_b}+\frac{1}{K_f}+\frac{1}{K_\psi}
\label{condition}
\end{equation}
defines the particular case where the scaling dimensions
$\mathrm{dim}[G_\phi]$ and $\mathrm{dim}[G_\theta]$
become identical and the factor $A_1$ in the
second terms of the right-hand side of Eqs.\ (\ref{eq:RG-2}) and (\ref{eq:RG-3})
vanishes.
Thus, Eq.\ (\ref{condition}) determines the phase boundary between
the phase where the $G_\phi$ operator is relevant
and the phase where the $G_\theta$ operator is relevant.
In the case where $K_f+K_\psi<K_b^{-1}+K_f^{-1}+K_\psi^{-1}$,
the coupling $G_\phi$ is relevant and renormalized to strong
coupling with $G_\phi>0$.
We note that the positive $G_\phi$ coupling implies
\textit{repulsive} density-density interactions 
between $f$ and $\psi$ particles.
On the other hand,
in the opposite case where $K_f+K_\psi<K_b^{-1}+K_f^{-1}+K_\psi^{-1}$,
the coupling $G_\theta$ is relevant and renormalized to strong coupling
with $G_\theta<0$.

The resulting phase diagram is shown in Fig.\ \ref{fig:pd}(b),
for which
the nature of the ground state in each phase is discussed in the
next section.

\vspace*{-.3cm}

\section{Phase diagram in the commensurate case}\label{sec:5}

\vspace*{-.3cm}

In the preceding section we determined the phase boundaries in the
phase diagram that admit quantum phase transitions. Therefore, in a
given region of relevance where a particular phase variable is locked,
the properties of the resulting phase that may exhibit dominant QLRO
can be understood by analyzing the exponents of the order parameter
correlations. 

For this purpose, the analysis based on the RG equations given by Eqs.\
(\ref{eq:RG}) is not simple since the $G_{3p}$ term contains
 both $\phi_s$ and $\theta_s$ ($s=f,\psi$) fields.
When treating this type of term, one often encounters subtleties in determining
ground-state phases, especially in the case that the $G_{3p}$
 term becomes relevant.
Thus it is necessary to make transformation
to a suitable basis.

\vspace*{-.3cm}

\subsection{Recombination of phase variables}

\vspace*{-.3cm}

We perform the following canonical transformation:
\begin{equation}
\bm \varphi  (x) = P \, \bm \phi  (x),
\qquad
\bm \vartheta  (x) = Q \, \bm \theta  (x),
\label{eq:nonunitarytrans}
\end{equation}
where
\begin{subequations}
\begin{eqnarray}
 \bm \phi  (x) \!\!&=&\!\! \left( 
\begin{array}{*{20}{c}}
   \phi_b(x)  \\
   \phi_f(x)  \\
   \phi_\psi(x)  
\end{array}
\right),
\qquad
\bm \theta  (x) = \left( 
\begin{array}{*{20}{c}}
   \theta_b(x)  \\
   \theta_f(x)  \\
   \theta_\psi(x)  
\end{array}
\right), \\ 
 \bm \varphi (x) \!\!&=&\!\! \left( 
\begin{array}{*{20}{c}}
   \varphi_1(x)  \\
   \varphi_2(x)  \\
   \varphi_3(x)  
\end{array} 
\right),
\qquad
\bm {\vartheta} (x) = \left( 
\begin{array}{*{20}{c}}
   \vartheta_1(x)  \\
   \vartheta_2(x)  \\
   \vartheta_3(x)
\end{array} 
\right).\quad
\end{eqnarray}
\end{subequations}
The transformation matrices
$P$ and $Q$ are generally nonorthogonal,
but the commutation relations of
$\varphi$ and $\vartheta$,
$[\varphi_{a}(x),\pi^{-1}\partial_{y}\vartheta_b(y)]
 =i\delta_{a,b}\delta(x-y)$,
are preserved as long as the relation $P Q^\mathrm{T}=1$ is satisfied
 \cite{Muttalib1986}.
A simplification of Eq.\ (\ref{eq:Hg3p}) follows from the following
choice of the matrices:
\begin{equation}
P = 
\frac{1}{\sqrt{2}}\left(
\begin{array}{rrr}
-2 & ~~1 & -1
\\
0 & 1 & 1
\\
0 & 1 & -1
\end{array}
\right), \,\,
Q = 
\frac{1}{\sqrt{2}}
\left(
\begin{array}{rrr}
-1 & ~~0 & 0
\\
0 & 1 & 1
\\
1 & 1 & -1
\end{array}
\right).
\label{eq:matrixPQ}
\end{equation}
Substituting the phase variables $\bm{\varphi}$ and $\bm{\vartheta}$,
we rewrite the cosine terms in Eqs.\ (\ref{eq:action}) as
\begin{subequations}
\begin{eqnarray}
S_{I,1}\!\!&=&\!\! 
\frac{G_{3p}}{i\pi}
\int \frac{d^2 r}{\alpha^2}
\cos(\sqrt{2}\vartheta_3)
\sin(\sqrt{2}\varphi_3) , \qquad
\\
S_{I,2}\!\!&=&\!\! 
\frac{G_{\phi}}{\pi}
\int \frac{d^2 r}{\alpha^2}
\cos(2\sqrt{2}\varphi_3) ,
\\
S_{I,3}\!\!&=&\!\! 
\frac{G_{\theta}}{\pi}
\int \frac{d^2 r}{\alpha^2}
\cos(2\sqrt{2}\vartheta_3).
\end{eqnarray}%
\label{eq:Hg3p-rotated}%
\end{subequations}
We note that the phase variables $\varphi_3$ and $\vartheta_3$
are under the influence of the $G_\phi$ and $G_\theta$ cosine
potentials, respectively.
In terms of the phase variables $\bm{\varphi}$ and $\bm{\vartheta}$,
the TLL Hamiltonian (\ref{eq:H0}) is rewritten as
\begin{eqnarray}
H_0\!\!&=&\!\!
H_b+H_f+H_\psi \nonumber\\
&=&\!\!
\frac{1}{2\pi}
\int {dx} \left[
 (\partial_x \bm \varphi^\mathrm{T}) M (\partial_x \bm \varphi) 
+(\partial_x \bm \vartheta^\mathrm{T}) N (\partial_x \bm \vartheta) 
\right],
\nonumber\\&&
\label{eq:H0-rotated}
\end{eqnarray}
where $M$ and $N$ are real symmetric matrices defined by
\begin{widetext}
\begin{subequations}
\begin{eqnarray}
M\!\!&=&\!\!
\frac{1}{2}\!
\left(
\begin{array}{ccc}
u_bK_b^{-1}
& 0 
& -u_bK_b^{-1}
\\
0
&\quad u_fK_f^{-1}+u_\psi K_\psi^{-1}
& u_fK_f^{-1}-u_\psi K_\psi^{-1}
\\ 
-u_bK_b^{-1}
&\quad u_fK_f^{-1}-u_\psi K_\psi^{-1}
&\quad u_bK_b^{-1}+u_fK_f^{-1}+u_\psi K_\psi^{-1}
\end{array}
\right),
\\
N\!\!&=&\!\!
\frac{1}{2}\!
\left(
\begin{array}{ccc}
4 u_bK_b +u_f K_f + u_\psi K_\psi
& \quad
u_f K_f - u_\psi K_\psi
& \quad
u_f K_f + u_\psi K_\psi
\\
u_f K_f - u_\psi K_\psi
& \quad
u_f K_f + u_\psi K_\psi
& \quad
u_f K_f - u_\psi K_\psi
\\ 
u_f K_f + u_\psi K_\psi
&\quad
u_f K_f - u_\psi K_\psi
& \quad
u_f K_f + u_\psi K_\psi
\end{array}
\right).
\end{eqnarray}%
\label{eq:matrixNM}%
\end{subequations}
\end{widetext}

The order parameters introduced in Sec.\ \ref{section:OP} 
can now be expressed in terms of the new phase variables
$\bm\varphi$ and $\bm\vartheta$.
The order parameters for the $b$ particles are given by
\begin{subequations}
\begin{eqnarray}
\mathcal{O}^{\mathrm{SF}}_b (x)
\!\! &\simeq& \!\!
e^{-i\sqrt{2}\vartheta_1} ,
\\
\mathcal{O}^{\mathrm{DW}}_b (x)
\!\! &\simeq& \!\!
e^{i2k_F^bx+i\sqrt{2}\varphi_1-i\sqrt{2}\varphi_3} .
\end{eqnarray}
The order parameters for the $p$-wave-pairing SF and 
out-of-phase DW states of the $f$ and $\psi$ particles are
written as
\begin{eqnarray}
\mathcal{O}^{\mathrm{SF}}_{f\psi} (x)
\!\! &\simeq& \!\!
e^{i\sqrt{2}\vartheta_2} 
 \sin (\sqrt{2}\varphi_3),
\\
\mathcal{O}^{\mathrm{DW}}_{f\psi} (x)
\!\! &\simeq& \!\! 
 e^{i2k_Fx -i\sqrt{2}\varphi_2} 
 \sin (\sqrt{2}\varphi_3),
\end{eqnarray}
from which it follows that
correlations of SF$_{f\psi}$ and DW$_{f\psi}$ 
are enhanced when the phase field $\varphi_3$ is locked at 
$\langle \sqrt{2}\varphi_3 \rangle = \pi/2 \, \mod\pi$.
Finally, the order parameters for the composite particles are
expressed as
\begin{eqnarray}
\mathcal{O}^{\mathrm{SF}}_{bff+b^\dagger\psi\psi} (x)
\!\! &\simeq& \!\! 
e^{i\sqrt{2}\vartheta_2}
\cos (\sqrt{2}\vartheta_3),
\\ 
\mathcal{O}^\mathrm{ph1}_{b^\dagger f^\dagger \psi}  (x)
\!\! &\simeq& \!\!
 e^{-i\sqrt{2}\varphi_3}
 \sin (\sqrt{2}\vartheta_3)
\nonumber \\ && \!{}
+
 e^{-i2k_F^b x-i\sqrt{2}\varphi_1}
 \cos(\sqrt{2}\vartheta_3), \quad
\label{eq:Oph1-2}
\\ 
\mathcal{O}^\mathrm{ph2}_{b^\dagger f^\dagger \psi}  (x)
\!\! &\simeq& \!\!
e^{i2k_F x -i\sqrt{2}\varphi_2 }
\cos (\sqrt{2}\vartheta_3).
\end{eqnarray}
\end{subequations}
We see that the correlations
of these order parameters are enhanced when 
the phase field $\vartheta_3$ is locked at 
$\langle \sqrt{2}\vartheta_3 \rangle = 0 \, \mod \pi$, except for the 
first contribution in Eq.\ (\ref{eq:Oph1-2}).

\vspace*{-.3cm}

\subsection{Effective low-energy Hamiltonian}

\vspace*{-.3cm}

The sinusoidal potentials of Eq.\ (\ref{eq:Hg3p-rotated})
take on forms similar to those of the spinless 
two-coupled chain system \cite{nersesyanPLA1993,yakovenkoJETP1992}
(see Appendix \ref{sec:app-A}).
In the two-chain system, operators generated in RG transformations
become relevant in the low-energy limit.
Similarly, we expect that
either the $G_\phi$ or $G_\theta$ term
can become relevant and renormalized to strong coupling,
as we have discussed below Eq.\ (\ref{condition}).
The relevant $G_\phi>0$ leads to locking of the phase field $\varphi_3$
at $\langle \sqrt{2} \varphi_3 \rangle = \pi/2 \mod \pi$,
whereas the relevant $G_\theta<0$ leads to the locking of the phase
field $\vartheta_3$ at
$\langle \sqrt{2} \vartheta_3 \rangle = 0 \mod \pi$.
When either $\varphi_3$ or $\vartheta_3$ is locked,
the remaining phase fields $\varphi_s$ and $\vartheta_s$ ($s=1,2$)
remain gapless, and then the system is effectively described by
a two-component TLL and a massive sine-Gordon model.
However, in contrast to the simple forms of sinusoidal potentials, 
the quadratic Hamiltonian in Eq.\ (\ref{eq:H0-rotated}) is complicated
by the presence of many cross terms.
One approach that we will implement here is  
to integrate out the massive mode $(\varphi_3,\vartheta_3)$
in a manner similar to
Ref.\ \cite{Ledermann:2000wo}, thereby
reducing the problem to a two-band system which can be exactly
diagonalized. 
To be more precise, 
when $G_\phi(l)\to + \infty$ in the RG analysis, 
the quantum fluctuations of the $\varphi_3$ field are suppressed,
and we can make the approximation
$\partial_x \varphi_3 \to \partial_x \langle \varphi_3\rangle \sim 0$. 
Moreover, since the cosine potentials can be ignored for the
strongly fluctuating $\vartheta_3$ field,
$\vartheta_3$ can be integrated out by completing the square for
$\partial_x \vartheta_3$ in the quadratic Hamiltonian,
as described in Ref.\ \cite{Ledermann:2000wo}.
The same approach can be used for $G_\theta(l)\to - \infty$. 
Consequently, the system can be described effectively by
the two-component TL liquid with the effective low-energy Hamiltonian
\begin{equation}
H^{\mathrm{eff}}
=
\sum_{i,j=1,2}
\int \frac{dx}{2\pi} \!\left(
\bar M_{ij}\varphi'_i\varphi'_j
+\bar N_{ij}\vartheta'_i\vartheta'_j
\right) ,
\label{eq:Heff}
\end{equation}
where $\varphi'_i=\partial_x\varphi_i$ and
$\vartheta'_i=\partial_x\vartheta_i$.
In the case when $\varphi_3$ is locked ($G_\phi\to\infty$), 
the renormalized coefficients are given by
$\bar M_{ij}=M_{ij}$ and
$\bar N_{ij}=N_{ij}-N_{i3}N_{j3}/N_{33}$ ($i,j=1,2$).
Similarly, when $\vartheta_3$ is locked ($G_\theta\to-\infty$), 
the coefficients are given by $\bar M_{ij}=M_{ij}-M_{i3}M_{j3}/M_{33}$ 
and $\bar N_{ij}=N_{ij}$.

The Hamiltonian (\ref{eq:Heff}) can be diagonalized
sequentially \cite{Muttalib1986}, yielding
\begin{eqnarray}
H^\mathrm{eff}
\!\!&=&\!\!
\frac{u_1}{2\pi}
\int {dx} \left[
  (\partial_x  \tilde \varphi_1)^2
+ (\partial_x  \tilde \vartheta_1)^2
\right]
\nonumber \\
&& {}
+
\frac{u_2}{2\pi}
\int {dx} \left[
  (\partial_x  \tilde \varphi_2)^2 
+ (\partial_x  \tilde \vartheta_2)^2
\right].
\label{eq:H^eff}
\end{eqnarray}
The canonical transformation between the phase variables 
$(\varphi, \vartheta)$ and $(\tilde\varphi,\tilde\vartheta)$ are given by
\begin{equation}
\left(
\begin{array}{c}
\varphi_1 \\ \varphi_2
\end{array}
\right)
=
\bar P
\left(
\begin{array}{c}
\tilde \varphi_1 \\ \tilde \varphi_2
\end{array}
\right),
\quad 
\left(
\begin{array}{c}
\vartheta_1 \\ \vartheta_2
\end{array}
\right)
=
\bar Q
\left(
\begin{array}{c}
\tilde \vartheta_1 \\ \tilde \vartheta_2
\end{array}
\right),
\end{equation}
where the transformation matrices $\bar P$ and $\bar Q$ are defined as
$\bar P = R_1 \Delta_1^{-1/2} R_2 \Delta_2^{1/4}$
and
$\bar Q = R_1 \Delta_1^{1/2} R_2 \Delta_2^{-1/4}$
with $\Delta_1$ and $\Delta_2$ being diagonal matrices.
Here the rotation matrix $R_1$ diagonalizes the matrix $\bar M$
as $R_1^\mathrm{T}\bar M R_1=\Delta_1$, and
the rotation matrix $R_2$ diagonalizes the matrix 
$\Delta_1^{1/2} R_1^\mathrm{T} \bar N R_1 \Delta_1^{1/2}
=R_2\Delta_2R_2^\mathrm{T}$.
The velocities $u_1$ and $u_2$ are diagonal elements of $\Delta_2^{1/2}$.

\vspace*{-.3cm}

\subsection{Correlation exponents}

\vspace*{-.3cm}

In this section we calculate correlation exponents for order
parameters characterizing the phases in Fig.~\ref{fig:pd}(b).

For the Gaussian model (\ref{eq:H^eff}),
the correlation functions of vertex operators, $\exp(i\lambda_i\varphi_i)$
and $\exp(i\lambda_i\vartheta_i)$ with real parameters $\lambda_{1,2}$, 
show power-law decay,
\begin{subequations}
\begin{eqnarray}
&&
\langle e^{i\lambda_1 \varphi_1(x) + i \lambda_2 \varphi_2(x)} 
             e^{-i\lambda_1 \varphi_1(0)-i \lambda_2 \varphi_2(0)}\rangle
\nonumber \\
&& \qquad\qquad\qquad
\sim x^{-\frac{1}{2}\lambda_1^2 \eta_{\varphi1} 
          -\frac{1}{2}\lambda_2^2 \eta_{\varphi2} 
          - \lambda_1 \lambda_2 \eta_{\varphi12}
}, 
\qquad\qquad
\\
&&
\langle e^{i\lambda_1 \vartheta_1(x) + i \lambda_2 \vartheta_2(x)} 
             e^{-i\lambda_1 \vartheta_1(0)-i \lambda_2 \vartheta_2(0)} 
\rangle
\nonumber \\
&& \qquad\qquad\qquad
\sim x^{-\frac{1}{2}\lambda_1^2 \eta_{\vartheta1} 
          -\frac{1}{2}\lambda_2^2 \eta_{\vartheta2} 
          - \lambda_1 \lambda_2 \eta_{\vartheta12}
},\qquad\qquad
\end{eqnarray}
\end{subequations}
where the exponents are given by 
\begin{subequations}
\begin{eqnarray}
&&
\eta_{\varphi i} = \sum_{j=1,2}   \bar P_{ij}^2 , \quad
\eta_{\varphi12} = \sum_{j=1,2}  \bar P_{1j} \bar P_{2j} , \quad
\\
&&
\eta_{\vartheta i}= \sum_{j=1,2}  \bar Q_{ij}^2,  \quad
\eta_{\vartheta12}= \sum_{j=1,2}  \bar Q_{1j} \bar Q_{2j}.  \quad
\end{eqnarray}%
\label{eq:eta}%
\end{subequations}
These results can be applied to the cases of interest.

\vspace*{-.3cm}

\subsubsection{Case of relevant $G_\phi$}

\vspace*{-.3cm}

In the case when $G_\phi$ is renormalized to strong coupling
($G_\phi\to\infty$), 
the fluctuations in the $\vartheta_3$ field diverge, and, consequently,
the order parameters that contain the vertex operator of $\vartheta_3$
exhibit short-range correlations or exponential decay at large distances.
On the other hand, the locked field $\varphi_3$ can be replaced by
its average $\langle \sqrt{2}\varphi_3\rangle = \pi/2 \mod \pi$
in the order parameters that contain $\varphi_3$.
The correlation functions for the boson order parameters 
are then given by
\begin{subequations}
\begin{eqnarray}
\langle \mathcal{O}^\mathrm{SF}_b (x) \,
  \mathcal{O}_b^{\mathrm{SF}\dagger}(0) \rangle
&\sim& x^{-1/(2K_b)},
\\
\langle \mathcal{O}_b^\mathrm{DW}(x) \,
  \mathcal{O}_b^{\mathrm{DW}\dagger}(0) \rangle
&\sim& x^{-2K_b} e^{i2k_F^b x}. \qquad
\end{eqnarray}
\end{subequations}
We note that the exponents are unchanged
from those in the
$g_{3p}=0$ case [see Eqs.\ (\ref{eq:cor0b})] and that
the correlation functions of $b$ particles 
are controlled by the TLL parameter $K_b$.
(To be precise, $K_b$ should be replaced by its
renormalized value $K_b^*$, whose difference from $K_b$ is
on the order of $g_{3p}^4$.)
The SF$_b$ is dominant for $K_b>1/2$, while the DW$_b$ becomes dominant
for $K_b<1/2$.
For $f$ and $\psi$ particles,
slowly decaying correlation functions
are given by
\begin{subequations}
\begin{eqnarray}
\langle \mathcal{O}^{\mathrm{SF}}_{f\psi} (x) \,
 \mathcal{O}^\mathrm{SF\dagger}_{f\psi}(0) \rangle
&\sim& x^{-1/K_2},
\\
\langle \mathcal{O}_{f\psi}^\mathrm{DW} (x) \,
\mathcal{O}_{f\psi}^{\mathrm{DW}\dagger}(0) \rangle
&\sim& x^{-K_2}
e^{i2k_F x}, \qquad
\end{eqnarray}
where
\begin{equation}
K_2=
2\left[
\left(\frac{u_f}{K_f}+\frac{u_\psi}{K_\psi}\right)
\left(\frac{1}{u_fK_f}+\frac{1}{u_\psi K_\psi}\right)
\right]^{-1/2}.
\label{eq:corre_case1_exp}%
\end{equation}%
\label{eq:corre_case1}%
\end{subequations}
The most dominant order for $f$ and $\psi$ particles
is determined by $K_2$:
The SF$_{f\psi}$ state for $K_2>1$ 
and the DW$_{f\psi}$ state for $K_2<1$.
In the phase diagram shown in Fig.\ \ref{fig:pd}(b),
the region of relevant $G_\phi$ is classified into
four regions according to the most slowly decaying correlation
for the bosonic ($b$) and fermionic ($f$, $\psi$) particles.

Here we briefly discuss the correspondence to the results obtained
in Ref.\ \cite{Zhang:2005dr}, in which 
the $f$-$\psi$ paired state is predicted within a mean-field analysis
of a 3D model.
It is pointed out in Ref.\ \cite{Zhang:2005dr} that
the molecular conversion term induces a \textit{repulsive} density-density 
interaction between a fermionic atom and a molecule
through a lowest-order virtual process.
This effective interaction is consistent with the interaction vertex
$G_\phi>0$ generated in our perturbative RG analysis.
Furthermore, it is argued in Ref.\ \cite{Zhang:2005dr}
that, if the bosons are \textit{condensed}, the effective
interaction between a fermionic atom and a molecule can become
\textit{attractive}, thereby yielding the SF order of ``$s$-wave''
$f$-$\psi$ pairing state.
In the present 1D case, the mean-field theory is invalid
(bosons cannot condense), and the effective interaction $G_\phi$ is repulsive.
Therefore the $s$-wave $f$-$\psi$ pairing cannot be stabilized.
Instead, we obtain a ``$p$-wave'' $f$-$\psi$ pairing (or out-of-phase DW state 
of $f$ and $\psi$ particle) which can be stabilized due to 
the induced \textit{repulsive} interaction between $f$ and $\psi$ particles.
  
\vspace*{-.3cm}

\subsubsection{Case of relevant $G_\theta$}

\vspace*{-.3cm}

Next we consider the case where the phase field $\vartheta_3$ is locked.
The fluctuations of the $\varphi_3$ field are divergent, and
its order parameters exhibit short-range correlations.
The order parameters of our interest are those involving $\vartheta_3$,
which can be simplified by replacing $\sqrt{2}\vartheta_3$
with its expectation value 
$\langle \sqrt{2}\vartheta_3\rangle=0 \mod \pi$.
The correlation functions of these leading order parameters
exhibit algebraic decay,
\begin{subequations}
\begin{eqnarray}
\langle \mathcal{O}_b^{\mathrm{SF}} (x) \, 
\mathcal{O}_b^{\mathrm{SF}\dagger}(0) \rangle
\!\! &\sim& \!\! 
 x^{-\eta_{\vartheta 1}},
\\
\langle \mathcal{O}_{bff+b^\dagger\psi\psi}^{\mathrm{SF}} (x) \,
 \mathcal{O}_{bff+b^\dagger\psi\psi}^{\mathrm{SF}\dagger}(0) \rangle
\!\! &\sim& \!\! 
x^{-\eta_{\vartheta 2}}, \qquad
\label{eq:corre_bff}
\\
\langle \mathcal{O}_{b^\dag f^\dag \psi}^{\mathrm{ph1}} (x) \,
 \mathcal{O}_{b^\dag f^\dag \psi}^{\mathrm{ph1}\dag}(0) \rangle
\!\! &\sim& \!\! 
x^{-\eta_{\varphi 1}} e^{-i2k_F^b x} ,\quad\quad
\\
\langle \mathcal{O}_{b^\dag f^\dag \psi}^{\mathrm{ph2}} (x) \,
 \mathcal{O}_{b^\dag f^\dag \psi}^{\mathrm{ph2}\dag}(0) \rangle
\!\! &\sim& \!\! 
x^{-\eta_{\varphi 2}} e^{i2k_F x} .
\quad\qquad
\end{eqnarray}%
\label{eq:corre-theta}%
\end{subequations}
The correlation functions of the order parameters
$\mathcal{O}^\mathrm{SF}_{ff}(x)$ and 
$\mathcal{O}^\mathrm{SF}_{\psi\psi}(x)$
also exhibit algebraic decay.
However, these orders cannot dominate over those given
in Eqs.\ (\ref{eq:corre-theta}),
since their exponents
are always greater than those in Eqs.\ (\ref{eq:corre-theta}).

When $u_b=u_f=u_\psi$ and $K_f=K_\psi$, the 
Hamiltonian (\ref{eq:Heff}) takes a diagonal form, and
the exponents are simplified to
\begin{subequations}
\begin{eqnarray}
\eta_{\varphi1}=2K_b+K_f, \quad \eta_{\varphi2} = K_f, 
\\
\eta_{\vartheta1}=\frac{1}{2K_b+K_f} , \quad \eta_{\vartheta2} = \frac{1}{K_f}.
\end{eqnarray}%
\label{exponents}%
\end{subequations}%
In the parameter region in Fig.\ \ref{fig:pd}(b)
where $G_\theta$ flows to strong coupling,
the exponent $\eta_{\vartheta1}$ is always smaller than the others
in Eqs.\ (\ref{exponents}).
Hence, the SF$_b$ state is designated as the most dominant state.
We also note that the SF$_b$ correlation is enhanced as compared with
the case of $g_{3p}=0$ where $\eta_{\vartheta1}\to 1/(2K_b)$.

\vspace*{-.3cm}

\section{Discussion and concluding remarks}\label{sec:6}

\vspace*{-.3cm}

In summary, we have carried out a comprehensive study of a two-channel
Bose-Fermi mixture, for which the analysis and results presented here
can possibly be applied towards more general many-body problems
involving interacting multicomponent quantum liquids.

When the densities of the fermionic atoms and fermionic MB molecules
are identical, the Feshbach molecule conversion and disassociation, the
$g_{3p}$ term, can become relevant and induce an excitation gap,
while the system retains two gapless modes.
One appealing feature of the phase diagram in particular
 is the existence of a dominant composite
$p$-wave pairing state
$\Psi_f^L\Psi_\psi^R$, which occurs for fermions in both the open and
closed hyperfine channels, induced by an effective interaction
mediated by $b$ atoms.
Ultimately, we hope that the phase diagram
presented here should demonstrate more general features of composite
orders and indirect scattering processes that will manifest in higher
dimensions.

Although we have established the qualitative behavior of the phase
diagram for a wide range of interaction couplings, 
a better comparison with experiments will require
microscopic determination of the TLL parameters using numerical methods.
Since our model contains specific order parameters that
couple different atomic species, a direct experimental probe
must be sensitive to interspecies density correlations.
Time-of-flight spectroscopy is the most promising method,
as it can directly image an atomic cloud's density profile,
which should demonstrate specific commensurability in the presence
of density wavelike order \cite{blochrmp08}.
A possible experimental realization within the cold atoms systems
would involve a magnetic
trapping technique developed on atom chips \cite{Folman:2008tx}. 
Recently, the TLL signatures have been confirmed by observing certain
quasi-long-range order within the noise correlations between two
independent 1D bosonic atomic condensates created on an atom chip
\cite{Hofferberth:2008cw}.
As discussed in Ref.\ \cite{Mathey:2008bb}, the analysis of the
noise correlations would be also useful to detect the composite pairing
states proposed in the present paper, since this measurement would be
sensitive not only to density-wave fluctuations but also to pairing
fluctuations.

In order to make a proper comparison of the results obtained in this paper
with actual experiments in trapped cold atom systems,
we have to take into account 
the density inhomogeneity arising from the harmonic trap.
For this purpose, we can apply the 
local density approximation (LDA)
\cite{Molmer:1998wd}
when the range of the density variation
is much larger than the average interparticle distance.
In the incommensurate case ($\bar{\rho}_f \neq \bar{\rho}_\psi$),
where the system is described as the three-component TLL
in the homogeneous limit, 
the low-energy properties can be analyzed 
by the bosonization scheme based on the LDA
\cite{Recati:2003,Cazalilla_review}.
On the other hand, in the commensurate case
($\bar{\rho}_f = \bar{\rho}_\psi$), 
the extension of the RG analysis would  not be so straightforward.
The numerical studies on the trapped boson system in an optical lattice 
\cite{Batrouni:2002fs,Kollath:2004jc}
have shown 
the transition from a superfluid to a Mott insulating state 
in the so-called ``wedding cake'' structure with  
density plateaus of the Mott state,
which was indeed observed experimentally \cite{Folling:2006dn}.
Such a structure can be ascribed to the commensurability effect 
which is present 
when the number of bosons per site becomes integer.
Since the commensurability effects
can be represented as the sinusoidal potentials
in the bosonization scheme,
we expect that similar commensurate-incommensurate transitions 
should be realized when a trapping potential is taken into account
in the present system.

\vspace*{-.3cm}

\acknowledgments 

\vspace*{-.3cm}

We thank T.\ Giamarchi, E. Orignac, and Masahiro Sato for important
discussions.
S.A. acknowledges helpful conversations
with A.\ M.\ Tsvelik and support from the RIKEN FPR program.

\appendix

\section{Two-coupled chain revisited}\label{sec:app-A}

\vspace*{-.3cm}

The model which we consider in the present paper has a close connection to 
the model of spinless two chains coupled by the one-particle interchain hopping
\cite{nersesyanPLA1993,Orignac:1997vq}.
The model Hamiltonian for the two-coupled chains is given by
\begin{eqnarray}
H_{2\, \mathrm{chain}}\!\!&=&\!\!
\sum_{s=1,2} \int dx \,
iv
\left(\Psi^{L\dagger}_{s} \partial_x \Psi_{s}^L
-\Psi^{R\dagger}_{s} \partial_x \Psi_{s}^R\right)
\nonumber\\
&&{}
-t_\perp \sum_{p=L,R}\int dx 
\bigl(\Psi_{1}^{p\dagger} \Psi_{2}^p+\mathrm{h.c.} \bigr)
\nonumber\\
&&{}
+\!\int \! dx \,
\bigl[g \bigl(\rho_1 \, \rho_1 + \rho_2 \, \rho_2 \bigr)
+2 g' \rho_1  \rho_2\bigr],
\end{eqnarray}
where $p=L (R)$ refers to the left- (right-) moving particle and 
$s=1,2$ is the chain index. 
The couplings $g$ and $g'$ represent the intrachain and interchain 
interactions, respectively \cite{nersesyanPLA1993}.
In earlier works, the interchain hopping term is diagonalized by
introducing the bonding and antibonding band basis of the field operators,
and then the bosonization and RG methods are
applied to the field operators on the band basis
\cite{nersesyanPLA1993,Orignac:1997vq}.
In this appendix, we verify that the same results can be obtained 
by directly applying the 
bosonization to the field operators on the original chain basis.
The bosonized forms of the field operators are given by
\begin{equation}
\Psi_{s}^{L/R}(x) = \frac{\xi_s}{\sqrt{2\pi \alpha}}
e^{\mp ik_F x \pm i \phi_s(x) + i\theta_s(x) },
\end{equation}
where $s=1,2$ is the chain index and 
$\xi_s$ is the Klein factor satisfying $\xi_1\xi_2=i$.
The commutation relation of the phase variables is 
$[\phi_s(x),\theta_{s'}(x')]
= i\pi \delta_{s,s'} \Theta(-x+x')$.
Since a dominant phase can be determined by 
the locking position of $\phi_s$ or $\theta_s$,
we have to carefully apply the fusion rules for vertex operators.

With the symmetric and antisymmetric combinations of phase variables,
$\phi_\pm = (\phi_1 \pm \phi_2)/\sqrt{2}$ and
$\theta_\pm = (\theta_1 \pm \theta_2)/\sqrt{2}$,
the bosonized Hamiltonian is written as
\begin{eqnarray}
H_{2\, \mathrm{chain}}\!\!&=&\!\!
\frac{u_+}{2\pi}
\int dx
\left[
\frac{1}{K_+} (\partial_x \phi_+)^2 + K_+ (\partial_x \theta_+)^2
\right]
\nonumber \\ && {}
+
\frac{u_-}{2\pi}
\int dx
\left[
\frac{1}{K_-} (\partial_x \phi_-)^2 + K_- (\partial_x \theta_-)^2
\right]
\nonumber \\ && {}
+i\frac{u_- G_\perp}{\pi \alpha^2} 
\int dx
\cos \sqrt{2}\theta_-  \sin \sqrt{2}\phi_-
\nonumber \\ && {}
+ \frac{u_-\widetilde{G}_\phi}{\pi \alpha^2}\int dx \cos 2\sqrt{2}\phi_-
\nonumber \\ && {}
+ \frac{u_-\widetilde{G}_\theta}{\pi\alpha^2}\int dx \cos 2\sqrt{2}\theta_- ,
\label{eq:ap-H}
\end{eqnarray}
where 
$K_\pm \simeq 1- (g\pm g')/(\pi v)$, 
$u_\pm \simeq  v + (g \pm g')/\pi$, and
$G_\perp = 2t_\perp \alpha/u_-$.
The coupling constants $\widetilde{G}_\phi$ and $\widetilde{G}_\theta$
are initially zero 
but generated through the RG transformation.
Only the asymmetric fields $(\phi_-,\theta_-)$ are subject to the
sinusoidal potentials, and the symmetric fields $(\phi_+,\theta_+)$
remain free.

The order parameters characterizing the ground state are
written in the bosonized form as \cite{nersesyanPLA1993,Orignac:1997vq}
\begin{subequations}
\begin{eqnarray}
\!\!\!
O_{\mathrm{CDW}^\pi}(x)
\!\!&=&\!\!
\Psi_{1}^{L\dagger} \Psi_{1}^{R} - \Psi_{2}^{L\dagger} \Psi_{2}^R
\simeq 
e^{i2k_Fx - i \sqrt{2}\phi_+}  \sin \sqrt{2}\phi_- , 
\hspace*{-.5cm}
\nonumber \\
\label{eq:2chain-O-CDW}
\\
O_{\mathrm{OAF}}(x)
\!\!&=&\!\!
\Psi^{L\dagger}_{1}\Psi_{2}^R-\Psi_{2}^{L\dagger} \Psi_{1}^R
\simeq
e^{i2k_Fx - i\sqrt{2} \phi_+} \cos\sqrt{2}\theta_- ,
\hspace*{-.5cm}
\nonumber \\
\label{eq:2chain-O-OAF}
\\
O_{\mathrm{SC}^d}(x)
\!\!&=&\!\!
\Psi_{1}^L\Psi_{2}^R+\Psi_{2}^L\Psi_{1}^R
\simeq
e^{i \sqrt{2}\theta_+} \sin \sqrt{2}\phi_- ,
\label{eq:2chain-O-SCd}
\\
O_{\mathrm{SC}^s}(x)
\!\!&=&\!\!
\Psi_{1}^L\Psi_{1}^R+\Psi_{2}^L\Psi_{2}^R
\simeq
e^{i \sqrt{2}\theta_+}  \cos \sqrt{2}\theta_- ,
\label{eq:2chain-O-SCs}
\end{eqnarray}
\end{subequations}
where CDW$^\pi$, OAF, SC$^d$, and SC$^s$ stand for
charge-density wave, orbital antiferromagnetic, $d$-wave
superconducting, and $s$-wave superconducting states, respectively.

In order to analyze the low-energy behavior of the $\phi_-$ mode,
we apply the momentum-shell renormalization-group method \cite{giambook}.
First, we split the phase variable 
as $\phi_s=\phi_s'+h_s$ and $\theta_s=\theta_s'+\tilde{h}_s$,
where $\phi_s'$ and $h_s$ are the phase fields containing
low-momentum and high-momentum components, respectively,
\begin{subequations}
\begin{eqnarray}
\phi'_s(\bm r) &=& \int_{|k| \lesssim\Lambda'}
  \frac{d^2 k}{(2\pi)^2} \, e^{i\bm k\cdot \bm r} \, \phi_s(\bm k),
\\
h_s(\bm r) &=& \int_{\Lambda'\lesssim |k| \lesssim\Lambda}
  \frac{d^2 k}{(2\pi)^2} \, e^{i\bm k\cdot \bm r} \, \phi_s(\bm k),
\end{eqnarray}
\end{subequations}
where $\bm r=(x,u_s\tau)$, $\bm k=(k,\omega/u_s)$, 
$\bm{k}\cdot\bm{r}=kx-\omega \tau$ \cite{giambook},
and 
$\phi_s(\bm k)$ is the Fourier transform of
$\phi_s(x,\tau)$.
The fields $\theta'_s$ and $\tilde h_s$ are defined
similarly as low- and high-momentum components of
the conjugate fields $\theta_s$.
The RG equations are derived by integrating out the $h$ and $\tilde h$
fields with the help of Eq.\ (\ref{eq:hcorre}).

We perform RG transformations of the action $S$
by treating the interchain hopping part,
\begin{equation}
S_{\perp}
=
i\frac{G_\perp}{\pi}
\int \frac{d^2 r}{\alpha^2} \,
\cos \sqrt{2} \theta_-(\bm r) \,
\sin \sqrt{2} \phi_-(\bm r),
\end{equation}
as a weak perturbation, where $d^2r = u_- dxd\tau$.
In doing so, we have to pay special attention 
to the commutative properties.
The equal-time commutation relation between $\phi_s$ and $\theta_s$ 
is given by
\begin{equation}
[\phi_s(x),\theta_{s'}(x')]= i\pi \delta_{s,s'}\Theta(-x+x'),
\label{eq:commutation}
\end{equation}
and their correlation functions are given by
\cite{giambook}
\begin{subequations}
\begin{eqnarray}
\langle \phi_s(\bm r) \theta_s(0) \rangle 
&=& 
\frac{1}{2} F_2(\bm r) +\frac{i\pi}{4}
,\\
\langle \theta_s(\bm r) \phi_s(0) \rangle 
&=& 
\frac{1}{2} F_2(\bm r) -\frac{i\pi}{4},
\end{eqnarray}%
\label{eq:correlationfn}%
\end{subequations}
where $F_2(\bm r)=-i\mathrm{Arg}(y_\alpha+ix)$ with
$y_\alpha=u_s\tau + \alpha \, \mathrm{sgn}(\tau)$.
The last terms  $\pm i\pi/4$ in Eqs.\ (\ref{eq:correlationfn})
are added in order to reproduce the 
commutation relation (\ref{eq:commutation}).
Integrating out the $h_s$ fields yields
the $O(t_\perp^2)$ contribution to the action $S$,
\begin{eqnarray}
 - \frac{1}{2}\langle  S_{\perp}^2 \rangle_h^c
\!\! &=& \!\!
-
\frac{G^2_{\perp}}{32\pi^2}
\sum_{\epsilon,\epsilon'=\pm}
\int \frac{d^2r_1}{\alpha^2} \frac{d^2r_2}{\alpha^2} 
\nonumber \\ && {}  \!\!\!\!\!\! \times \!\!
\langle
e^{i\epsilon \sqrt{2}\theta_-(1)}
e^{i\epsilon' \sqrt{2}\phi_-(1)}
e^{-i\epsilon \sqrt{2}\theta_-(2)}
e^{i\epsilon' \sqrt{2}\phi_-(2)}
\rangle_h^\mathrm{c}  \hspace*{-.5cm}
\nonumber \\ && {}  \hspace*{-.5cm}
+
\frac{G^2_{\perp}}{32\pi^2}
\sum_{\epsilon,\epsilon'=\pm}
\int \frac{d^2r_1}{\alpha^2} \frac{d^2r_2}{\alpha^2} 
\nonumber \\ && {}  \!\!\!\!\!\! \times \!\!
\langle
e^{i\epsilon \sqrt{2}\theta_-(1)}
e^{i\epsilon' \sqrt{2}\phi_-(1)}
e^{i\epsilon \sqrt{2}\theta_-(2)}
e^{-i\epsilon' \sqrt{2}\phi_-(2)}
\rangle_h^\mathrm{c} , \hspace*{-.5cm}
\nonumber \\
\end{eqnarray}
where $(1)$ and $(2)$ stand for $(\bm r_1)$ and 
$(\bm r_2)$, respectively, and
$\langle \cdots \rangle_h^\mathrm{c}$ is the cumulant expectation with
respect to the $h$ and $\tilde h$ fields.
The cumulant expectations can be evaluated as
\begin{eqnarray}
&&
\hspace*{-1cm}
\langle
e^{i\epsilon \sqrt{2}\theta_-(1)}
e^{i\epsilon' \sqrt{2}\phi_-(1)}
e^{-i\epsilon \sqrt{2}\theta_-(2)}
e^{i\epsilon' \sqrt{2}\phi_-(2)}
\rangle_h^\mathrm{c}
\nonumber \\
&=& 
e^{i\epsilon \sqrt{2}\theta_-'(1)}
e^{i\epsilon' \sqrt{2}\phi_-'(1)}
e^{-i\epsilon \sqrt{2}\theta_-'(2)}
e^{i\epsilon' \sqrt{2}\phi_-'(2)}
\nonumber \\ && {} \times
\,
e^{-(K_-+K_-^{-1})\delta g(\bm 0)}
\Bigl[  e^{(K_-^{-1}-K_-)\delta g(r_{12})} - 1  \Bigr].
\quad \,\,
\end{eqnarray}
We note that the integrand becomes nonzero only 
for small $r_{12}/\alpha=|\bm r_1-\bm r_2|/\alpha$ since the function
$\delta g(r)$ decays rapidly in $r/\alpha$.
We can rewrite the product of the vertex
operators  as 
\begin{eqnarray}
&& \hspace*{-1cm}
e^{i\epsilon \sqrt{2}\theta_-'(1)}
e^{i\epsilon' \sqrt{2}\phi_-'(1)}
e^{-i\epsilon \sqrt{2}\theta_-'(2)}
e^{i\epsilon' \sqrt{2}\phi_-'(2)}
\nonumber \\ 
&\approx&
-e^{i\epsilon \sqrt{2}\theta_-'(1)
  +i\epsilon' \sqrt{2}\phi_-'(1)
  -i\epsilon \sqrt{2}\theta_-'(2)
  +i\epsilon' \sqrt{2}\phi_-'(2)},
\quad \,\,
\label{eq:product}
\end{eqnarray}
where we have used 
$[\phi'_s(\bm r_1) , \theta'_s(\bm r_2)]
 + [\phi'_s(\bm r_2) , \theta'_s(\bm r_1)]
=
 \langle [\phi'_s(\bm r_1) , \theta'_s(\bm r_2)] \rangle
 + \langle [\phi'_s(\bm r_2) , \theta'_s(\bm r_1)]\rangle  = i\pi$,
together with the relation (\ref{eq:correlationfn}).
Using Eq.\ (\ref{eq:ope}),
we can perform the operator-product expansion 
[$r_{12}=|\bm r_1- \bm r_2|$ and 
$\bm R=(\bm r_1 + \bm r_2)/2$]:
\begin{eqnarray}
&& {} \hspace*{-.7cm}
e^{i\epsilon \sqrt{2}\theta_-'(1)
  +i\epsilon' \sqrt{2}\phi_-'(1)
  -i\epsilon \sqrt{2}\theta_-'(2)
  +i\epsilon' \sqrt{2}\phi_-'(2)}
\nonumber \\
&\approx&
\, :
e^{i\epsilon \sqrt{2}\theta_-'(1)
  +i\epsilon' \sqrt{2}\phi_-'(1)
  -i\epsilon \sqrt{2}\theta_-'(2)
  +i\epsilon' \sqrt{2}\phi_-'(2)} : \,
\nonumber \\ && {} \times
e^{-K_-^{-1}[\bar g(0)-\bar g(r_{12})]}
e^{-K_-[\bar g(0)+\bar g(r_{12})]}
\nonumber \\
&\approx&
\, :
e^{ +i\epsilon' 2\sqrt{2}\phi_-'(\bm R) } : \,
\,
e^{-K_-^{-1}[\bar g(0)-\bar g(r_{12})]}
e^{-K_-[\bar g(0)+\bar g(r_{12})]}
\nonumber \\
&=&
e^{ +i\epsilon' 2\sqrt{2}\phi_-'(\bm R) } 
e^{-(K_-^{-1}-K_-)[\bar g(0)-\bar g(r_{12})]},
\end{eqnarray}
where
$\bar g(r)$ is given in Eq.\ (\ref{eq:phipcorre}),
and we have used Eqs.\ (\ref{eq:correlationfn}).
Thus we find
\begin{eqnarray}
 - \frac{1}{2}\langle  S_{\perp}^2 \rangle_h^\mathrm{c}
\!\! &=& \!\!
+\frac{G^2_{\perp}}{4\pi}
e^{-(K_-+K_-^{-1})\delta g(\bm 0)}
A_1\biglb((K_-^{-1}-K_-)/2\bigrb)
\, dl \!\!\!\!
\nonumber \\ && {} \quad \times
\int \frac{d^2R}{\alpha^2} 
\cos 2\sqrt{2}\phi_-'(\bm R) 
\nonumber \\ && {}  \!\!\!\!
-\frac{G^2_{\perp}}{4\pi}
e^{-(K_-+K_-^{-1})\delta g(\bm 0)}
A_1\biglb((K_--K_-^{-1})/2\bigrb)
\, dl \!\!\!\!
\nonumber \\ && {} \quad \times
\int \frac{d^2R}{\alpha^2} 
\cos 2\sqrt{2}\theta_-'(\bm R) ,
\end{eqnarray}
where $A_i(\beta)$ is defined in Eqs.\ (\ref{eq:A}).
The first (second) term renormalizes the $\tilde G_\phi$ ($\tilde
G_\theta$) term.
The full RG equations for the coupling constants and
the TLL parameter $K_-$ are given by
\begin{subequations}
\begin{eqnarray}
\frac{dG_\perp}{dl} \!\!&=&\!\!
\left(2-\frac{K_-}{2}-\frac{1}{2K_-}\right)G_\perp,
\\
\frac{d\widetilde{G}_\phi}{dl} \!\! &=& \!\!
(2-2K_-) \widetilde{G}_\phi 
+ \frac{1}{4} G_\perp^2 A_1 \biglb((K_-^{-1}-K_-)/2\bigrb),
\label{dG_phi/dl}
\nonumber \\
\\
\frac{d\widetilde{G}_\theta}{dl}  \!\! &=& \!\!
(2-2K_-^{-1}) \widetilde{G}_\theta 
- \frac{1}{4} G_\perp^2 A_1\biglb((K_--K_-^{-1})/2\bigrb) ,
\label{dG_theta/dl}
\nonumber \\
\\
\frac{dK_-}{dl}
\!\! &=& \!\! 
-2\widetilde{G}_\phi^2 K_-^2 A_2\!\left(2K_-\right)
+ 2\widetilde{G}_\theta^2  A_2\!\left(2K_-^{-1}\right).
\end{eqnarray}%
\label{eq:ap-RG}%
\end{subequations}

We see from Eqs.\ (\ref{dG_phi/dl}) and (\ref{dG_theta/dl}) that
the one-loop RG processes yield contributions of order $G_\perp^2$
to $\widetilde{G}_\phi$ and $\widetilde{G}_\theta$, respectively.
Consequently, when $K_-<1$, the coupling $\widetilde{G}_\phi$ is relevant
and renormalized to strong coupling ($\widetilde{G}_\phi\to + \infty$).
In this case, the phase field $\phi_-$ is locked at 
$ \langle\sqrt{2}\phi_-\rangle=  \pi/2  \mod \pi$.
On the other hand, if $K_->1$, the coupling $\widetilde{G}_\theta$ 
is relevant and renormalized to strong coupling
($\widetilde{G}_\theta\to -  \infty$),
and then the phase field $\theta_-$
is locked at 
$\langle\sqrt{2}\theta_-\rangle=  0  \mod \pi$.

For $K_-<1$ (i.e., $g>g'$),
the relevant order parameters, CDW$^\pi$ and SC$^d$, are reduced to
$O_{\mathrm{CDW}^\pi}(x)\to e^{i2k_Fx - i \sqrt{2}\phi_+}$, 
$O_{\mathrm{SC}^d}(x) \to e^{i \sqrt{2}\theta_+} $, as $\phi_-$ is
locked at $\langle\sqrt2\phi_-\rangle=\pi/2 \mod \pi$.
These correlation functions show QLRO,
\begin{subequations}
\begin{eqnarray}
\langle O_{\mathrm{CDW}^\pi}(x)
 O_{\mathrm{CDW}^\pi}^\dagger(0)\rangle
\!&\sim&\! x^{-K_+} e^{i2k_Fx},
\\
\langle O_{\mathrm{SC}^d}(x) O_{\mathrm{SC}^d}^\dagger(0)\rangle
\!&\sim&\! x^{-1/K_+}.
\end{eqnarray}
\end{subequations}
The dominant correlation is determined by the value of
the TLL parameter $K_+$;
the CDW (SC$^d$) state becomes most dominant for $K_+<1$ ($K_+>1$),
i.e., $g+g'>0$ ($g+g'<0$).

For $K_->1$ (i.e., $g<g'$), 
the relevant order parameters are given by
$O_{\mathrm{OAF}}(x)\to e^{i2k_Fx - i\sqrt{2} \phi_+}$ and
$O_{\mathrm{SC}^s}(x)\to e^{i \sqrt{2}\theta_+} $,
and their correlation functions are
\begin{subequations}
\begin{eqnarray}
\langle O_{\mathrm{OAF}}(x) O_{\mathrm{OAF}}^\dagger(0) \rangle
\!&\sim&\! x^{-K_+} e^{i2k_Fx},\\
\langle O_{\mathrm{SC}^s}(x) O_{\mathrm{SC}^s}^\dagger(0)\rangle
\!&\sim&\! x^{-1/K_+}.
\end{eqnarray}
\end{subequations}
The dominant correlation is 
the OAF (SC$^s$) state when  $K_+<1$ ($K_+>1$),
i.e., $g+g'>0$ ($g+g'<0$).

Since the RG analysis described above correctly reproduces
the phase diagram obtained in Ref.\ \cite{Orignac:1997vq},
the validity of our method is confirmed.
As we noted earlier, the sinusoidal potentials of 
the two-chain Hamiltonian (\ref{eq:ap-H}) have forms similar to those
of Eqs.\ (\ref{eq:Hg3p-rotated}).
We can thus study the phase diagram of our model
using the same RG method (with straightforward generalization),
as described in Secs.\ \ref{sec:RG} and \ref{sec:5}, 
where the RG equations (\ref{eq:RG}) are indeed similar to
Eqs.\ (\ref{eq:ap-RG}).

\vspace*{-.3cm}

\section{Mapping to two-coupled chain with gauge field}\label{sec:app-B}

\vspace*{-.3cm}

The model Hamiltonian (\ref{eq:fullham}) can be mapped to the
Hamiltonian for the two-coupled chain with gauge field.
By applying the phase representation only for the boson
[Eq.\ (\ref{eq:psi_b})], and by expressing $\Psi_f\to \Psi_1$ and 
$\Psi_\psi \to \Psi_2$, the effective Hamiltonian is expressed as
\begin{eqnarray}
H_{f\psi}
\!\! &\equiv& \!\!
H_f+H_\psi
\nonumber \\
\!\! &=& \!\!
\sum_{s=1,2} 
\int dx \,
iu
\Bigl(
 \Psi^{L\dagger}_{s} \partial_x \Psi^L_{s}
-\Psi^{R\dagger}_{s} \partial_x \Psi^R_{s}
\Bigr)
\nonumber \\ && {}
+\sum_{s=1,2} 
 g \int dx  \rho_s(x) \rho_s(x+\delta) ,
\\
H_b
\!\! &=& \!\! 
\frac{u}{2\pi}\int dx 
  \left[
    \frac{1}{K_b} \left( \partial_x \phi_b \right)^2
  + K_b \left( \partial_x \theta_b \right)^2
  \right],\quad
\\
H_{3p}
\!\! &=& \!\!
-t_\perp \sum_{p=L,R}
\int dx 
\left(\Psi_{1}^{p\dagger} \Psi_{2}^p
 e^{-i\theta_b}+\mathrm{h.c.}\right),
\qquad
\end{eqnarray}
where $t_\perp \equiv -g_{3p}/(2\pi\alpha)^{1/2}$,
and we have set $u_b=u_f=u_\psi \, (\equiv u)$.
We assumed the short-range interaction $V_{ff}(x-x')=g_f\delta(x-x'\pm \delta)$
and  $V_{\psi\psi}(x-x')=g_\psi\delta(x-x'\pm \delta)$, where $\delta$ is
the small quantity, and we set $g_f=g_\psi(\equiv g)$ for simplicity.
We focus on the commensurate case $\rho_f=\rho_\psi$.

This model can be interpreted as the two-coupled chain model with 
a gauge field on the \textit{rung}.
In this section, we first eliminate the effect of the gauge field 
by gauge transformation and diagonalize the $t_\perp$ term.
In the next step, we apply the bosonization, 
as performed in the two-chain problem \cite{Orignac:1997vq}.
By the gauge transformation
\begin{eqnarray}
\Psi_{1}^p(x) &\to& \tilde\Psi_{1}^p(x)= \Psi_{1}^p(x)\, e^{+ i\theta_b(x)/2},
\\
\Psi_{2}^p(x) &\to& \tilde\Psi_{2}^p(x)= \Psi_{2}^p(x)\, e^{- i\theta_b(x)/2},
\end{eqnarray}
the $t_\perp$ term becomes
$-t_\perp \sum_p \int dx (\tilde\Psi_{1}^{p\dagger} \tilde\Psi^p_{2}
 +\mathrm{H.c.})$.
Since the $t_\perp$ term is expressed in the form of
conventional interchain hopping, we can follow the approach
of Ref.\ \cite{Orignac:1997vq} 
in which the relevant $t_\perp$ term was treated nonperturbatively.

The interchain hopping term 
can be diagonalized by introducing the bonding and antibonding
operators:
\begin{equation}
\Psi_{+}^p
=
\frac{1}{\sqrt{2}}(\tilde\Psi_{1}^p + \tilde\Psi_{2}^p ),
\quad
\Psi_{-}^p
=
\frac{1}{\sqrt{2}}(\tilde\Psi_{1}^p - \tilde\Psi_{2}^p ).
\end{equation}
The $t_\perp$ term is given by 
$-t_\perp \sum_p \int dx 
( \Psi_{+}^{p\dagger} \Psi^p_{+}
-\Psi_{-}^{p\dagger} \Psi^p_{-})$, and the 
intrachain  kinetic terms  are given by
\begin{eqnarray}
&&
\int dx \,
iu \! \left(
\Psi^{L\dagger}_{+}
 \partial_x 
\Psi^L_{+}
-
\Psi^{R\dagger}_{+}
 \partial_x
\Psi^R_{+}
\right)
\nonumber \\ && {}
+
\int dx \,
iu \! \left(
\Psi^{L\dagger}_{-}
 \partial_x 
\Psi^L_{-}
-
\Psi^{R\dagger}_{-}
 \partial_x
\Psi^R_{-}
\right)
\nonumber \\ && {}
+ \frac{u}{2}
\int dx 
\left(
\Psi^{L\dagger}_{+} 
\Psi^L_{-} 
-
\Psi^{R\dagger}_{+}  
\Psi^R_{-}
+
\mathrm{H.c.}
\right)
( \partial_x  \theta_b) .
\quad
\label{eq:HfHpsi_kin}
\end{eqnarray}
In contrast to the $t_\perp$ term, which is given in a diagonalized form,
the intrachain  kinetic terms contain the gauge field and 
the field operators are given in the nondiagonalized form.

Now we bosonize the fields $\Psi_\pm$:
\begin{equation}
 \Psi^p_\pm  (x) 
=
\frac{\xi_\pm}{\sqrt{2\pi\alpha}}
e^{in[ {k_F x - \phi _\pm (x)} ] + i\theta_\pm (x)},
\end{equation}
where $n=+(-)$ for $p=R(L)$.
In order to simplify the notation, we further 
 apply the simple transformation,
\begin{equation}
\tilde \phi_+
=
\frac{1}{\sqrt{2}}
\left(
\phi_+ + \phi_-
\right)
,\quad
\tilde \phi_-
=
\frac{1}{\sqrt{2}}
\left(
\phi_+ - \phi_-
\right),
\end{equation}
and then the Hamiltonians $H_{f\psi}$ and $H_{3p}$ are  
 expressed as
\begin{eqnarray}
H_{f\psi}
\!\! &=& \!\!
 \frac{u_+}{2\pi}
\int dx \left[\frac{1}{K_+} (\partial_x \tilde\phi_+)^2
 + K_+ (\partial_x \tilde\theta_+)^2 \right]
\nonumber \\  && {} \!\!
+
\frac{u_-}{2\pi}
\int dx \left[ \frac{1}{K_-}(\partial_x \tilde\phi_-)^2
 + K_- (\partial_x \tilde\theta_-)^2 \right]
\nonumber \\  && {} \!\!
- \frac{ig_0}{\pi \alpha}
\int dx \,
(\partial_x  \theta_b)
\sin \sqrt{2}\tilde\theta_-
\cos \sqrt{2}\tilde\phi_-
\nonumber \\  && {} \!\!
+
\frac{1}{2 \pi^2 \alpha^2}   
\int dx 
\Bigl[
-g_\phi \cos 2\sqrt{2}\tilde\phi_-
+g_\theta \cos 2\sqrt{2}\tilde\theta_-
\Bigr],
\nonumber \\
H_{3p}
 \!\! &=&  \!\!
t_\perp
\int dx 
\frac{\sqrt{2}}{\pi}
\partial_x \tilde \phi_-,
\end{eqnarray}
where 
\begin{eqnarray}
g_0 =  u,
\quad
g_\phi = g_\theta 
= g  [1-\cos (2 k_F \delta )],
\end{eqnarray}
and $K_\pm$ and $u_\pm$ depend on $g$ and $u$.
The microscopic determination of the parameters $K_\pm$ and $u_\pm$ 
are given in Ref.\ \cite{Orignac:1997vq} for weak-coupling region,
while it requires numerical analysis in the wide range of interactions.

From the scaling analysis,
the RG equation for $g_0$ is given by 
$dg_0/dl=(1-K_-/2-K_-^{-1}/2)g_0$, implying that
the $g_0$ term is marginal for $K_-=1$ and becomes 
irrelevant for $K_-\neq 1$.
The presence of the marginal or irrelevant $g_0$ term would
give rise to slight renormalization of other quantities like
$K_b$, $K_-$, $g_\phi$, $g_\theta$, and $t_\perp$.
However, we argue below that, up to such relatively unimportant corrections,
we can safely neglect the $g_0$ term
so this Hamiltonian takes on the same form as two-coupled chains,
derived in Ref.\ \cite{Orignac:1997vq}.
First, we observe that
the $t_\perp$ term suppresses the potential $\cos 2\sqrt{2} \tilde\phi_-$,
since the former favors the incommensurate state while the latter favors
the commensurate state.
Depending on the value of $K_-$, the $\tilde\theta_-$ field can either 
remain massless or develop a gap.
We concentrate here on the case where $\tilde\theta_-$ develops a gap
and acquires a nonzero expectation value determined by minimizing the 
ground-state energy.
The average value of the massive field $\langle \sqrt{2}\tilde\theta_-\rangle$
depends on the sign of $g^*_\theta$, 
where $g^*_\theta$ is the fixed-point value of $g_\theta$.
The average value of $\langle \sqrt{2}\tilde\theta_-\rangle$ and 
the corresponding order parameters are summarized in Table~\ref{table:ap2}.

\begin{table}[t]
\caption{
The average value of $\langle\sqrt{2}\tilde\theta_-\rangle$ and
the corresponding order parameter,
 determined by the fixed point value of the relevant $g_\theta^*$.
}
\label{table:ap2}
\begin{ruledtabular}
\begin{tabular}{ccc}
Fixed point
&
Average value
&
Order parameter
\\
\hline
$g_\theta^* >0$ &
$\langle \sqrt{2}\tilde\theta_-\rangle
 = \pi/2 \,\, \mathrm{mod} \, \pi$
&
$\langle \sin\sqrt{2}\tilde \theta_- \rangle \neq 0$
\\
$g_\theta^* <0$ &
$\langle \sqrt{2}\tilde\theta_-\rangle
 = \displaystyle 0 \,\,\, \, \mathrm{mod} \, \pi$
&
$\langle \cos\sqrt{2}\tilde \theta_- \rangle \neq 0$
\\
\end{tabular}
\end{ruledtabular}
\end{table}

The order parameters of the interest are 
$\mathcal{O}^{\mathrm{DW}}_{f\psi}$ [Eq.\ (\ref{eq:O_DW_fpsi})], 
$\mathcal{O}_{f\psi}^\mathrm{SF}$ [Eq.\ (\ref{eq:fpsi})], and
$\mathcal{O}^{\mathrm{SF}}_{bff+b^\dagger\psi\psi}$ [Eq.\ (\ref{eq:Obff})].
After the gauge transformation and the bosonization,
these order parameters are written as
\begin{subequations}
\begin{eqnarray}
\mathcal{O}^{\mathrm{DW}}_{f\psi}(x)
&=&
\Psi_1^{L\dagger} \Psi_1^R - \Psi_2^{L\dagger} \Psi_2^R 
\nonumber \\
&=&
\tilde\Psi_1^{L\dagger} \tilde\Psi_1^R
   - \tilde\Psi_2^{L\dagger} \tilde\Psi_2^R
\nonumber \\
&=&
\Psi_+^{L\dagger} \Psi_-^R
   +  \Psi_-^{L\dagger} \Psi_+^R
\nonumber \\
&=&
\frac{1}{\pi \alpha} 
e^{+i2k_F x - i\sqrt{2} \tilde \phi_+ } 
\sin \sqrt{2}\tilde\theta_- ,
\\ 
\mathcal{O}_{f\psi}^\mathrm{SF} (x)
&=&
\Psi_1^L \Psi_2^R + \Psi_2^R \Psi_1^L 
\nonumber \\
&=&
\tilde\Psi_1^L \tilde\Psi_2^R + \tilde\Psi_2^L \tilde\Psi_1^R 
\nonumber \\
&=&
\Psi_+^L \Psi_+^R - \Psi_-^L \Psi_-^R 
\nonumber \\
&=&
\frac{1}{\pi\alpha}
e^{i\sqrt{2}\tilde\theta_+}
\sin \sqrt{2}\tilde\theta_- ,
\\ 
\mathcal{O}^{\mathrm{SF}}_{bff+b^\dagger\psi\psi}(x)
&=&  
\frac{1}{\sqrt{2\pi\alpha}}  
\left(
e^{i\theta_b} \, \Psi_1 ^L\Psi _1^R
+ 
e^{-i\theta_b} \, \Psi_2 ^L\Psi _2^R
\right)
 \nonumber \\
&=&  
\frac{1}{\sqrt{2\pi\alpha}}  \, 
\left(
\tilde\Psi _1 ^L \tilde\Psi_1^R
+ 
\tilde\Psi _2 ^L \tilde\Psi_2^R
\right)
 \nonumber \\
&=&  
\frac{1}{\sqrt{2\pi\alpha}}  \, 
\left(
\Psi_+^L \Psi_+^R
+ 
\Psi_-^L \Psi_-^R
\right)
\nonumber \\
&=&
\frac{-2i}{(2\pi\alpha)^{3/2}}
e^{i\sqrt{2}\tilde\theta_+}
\cos \sqrt{2}\tilde\theta_-,
\end{eqnarray}
\end{subequations}
where we have set $\xi_+\xi_-=i$.

When $g_{\theta}^*>0$ (see Table \ref{table:ap2}), 
we find that the correlation functions for 
$\mathcal{O}^{\mathrm{DW}}_{f\psi}$ and $\mathcal{O}_{f\psi}^\mathrm{SF}$
exhibit algebraic decay,
\begin{subequations}
\begin{eqnarray}
\langle \mathcal{O}^{\mathrm{SF}}_{f\psi} (x) \,
 \mathcal{O}^{\mathrm{SF}}_{f\psi}(0)^\dagger \rangle
\!\! &\sim& \!\! x^{-1/K_+},
\\
\langle \mathcal{O}^{\mathrm{DW}}_{f\psi} (x) \,
\mathcal{O}^{\mathrm{DW}}_{f\psi} (0)^\dagger  \rangle
\!\! &\sim& \!\! x^{-K_+}
e^{i2k_F x} 
.
\end{eqnarray}
\end{subequations}
This behavior is consistent with Eqs.\ (\ref{eq:corre_case1}) if 
we equate $K_+$ with $K_2$. 
We note that, in the simplified case where $u_b=u_f=u_\psi$ and $K_f=K_\psi$, 
the exponent $K_2$ is given by $K_2 \to K_f$,
as seen from Eq.\ (\ref{eq:corre_case1_exp}).

On the other hand, when $g_\theta^*<0$, 
the correlation function of
$\mathcal{O}^{\mathrm{SF}}_{bff+b^\dagger\psi\psi}$  exhibits
algebraic decay,
\begin{equation}
\langle \mathcal{O}^{\mathrm{SF}}_{bff+b^\dagger\psi\psi} (x) \,
 \mathcal{O}^{\mathrm{SF}}_{bff+b^\dagger\psi\psi}(0)^\dagger \rangle
\sim x^{-1/K_+} .
\end{equation}
This behavior is consistent with Eq.\ (\ref{eq:corre_bff})
by noting that
  the exponent $\eta_{\vartheta2}$ is  given by 
 $\eta_{\vartheta2}\to 1/K_f$ [Eqs.\ (\ref{exponents})]
in the case of $u_b=u_f=u_\psi$ and $K_f=K_\psi$.


\begin{thebibliography}{51}
\expandafter\ifx\csname natexlab\endcsname\relax\def\natexlab#1{#1}\fi
\expandafter\ifx\csname bibnamefont\endcsname\relax
  \def\bibnamefont#1{#1}\fi
\expandafter\ifx\csname bibfnamefont\endcsname\relax
  \def\bibfnamefont#1{#1}\fi
\expandafter\ifx\csname citenamefont\endcsname\relax
  \def\citenamefont#1{#1}\fi
\expandafter\ifx\csname url\endcsname\relax
  \def\url#1{\texttt{#1}}\fi
\expandafter\ifx\csname urlprefix\endcsname\relax\def\urlprefix{URL }\fi
\providecommand{\bibinfo}[2]{#2}
\providecommand{\eprint}[2][]{\url{#2}}

\bibitem{feshbach}
\bibinfo{author}{\bibfnamefont{H.}~\bibnamefont{Feshbach}},
  \bibinfo{journal}{Ann. Phys.} \textbf{\bibinfo{volume}{5}},
  \bibinfo{pages}{357 } (\bibinfo{year}{1958}).

\bibitem{jinPRL2004}
\bibinfo{author}{\bibfnamefont{C.~A.} \bibnamefont{Regal}},
  \bibinfo{author}{\bibfnamefont{M.}~\bibnamefont{Greiner}}, \bibnamefont{and}
  \bibinfo{author}{\bibfnamefont{D.~S.} \bibnamefont{Jin}},
  \bibinfo{journal}{Phys. Rev. Lett.} \textbf{\bibinfo{volume}{92}},
  \bibinfo{pages}{040403} (\bibinfo{year}{2004}).

\bibitem{ketterlePRL2004}
\bibinfo{author}{\bibfnamefont{M.~W.} \bibnamefont{Zwierlein}},
  \bibinfo{author}{\bibfnamefont{C.~A.} \bibnamefont{Stan}},
  \bibinfo{author}{\bibfnamefont{C.~H.} \bibnamefont{Schunck}},
  \bibinfo{author}{\bibfnamefont{S.~M.~F.} \bibnamefont{Raupach}},
  \bibinfo{author}{\bibfnamefont{A.~J.} \bibnamefont{Kerman}},
  \bibnamefont{and} \bibinfo{author}{\bibfnamefont{W.}~\bibnamefont{Ketterle}},
  \bibinfo{journal}{Phys. Rev. Lett.} \textbf{\bibinfo{volume}{92}},
  \bibinfo{pages}{120403} (\bibinfo{year}{2004}).

\bibitem{ThomasPRL2004}
\bibinfo{author}{\bibfnamefont{J.}~\bibnamefont{Kinast}},
  \bibinfo{author}{\bibfnamefont{S.~L.} \bibnamefont{Hemmer}},
  \bibinfo{author}{\bibfnamefont{M.~E.} \bibnamefont{Gehm}},
  \bibinfo{author}{\bibfnamefont{A.}~\bibnamefont{Turlapov}}, \bibnamefont{and}
  \bibinfo{author}{\bibfnamefont{J.~E.} \bibnamefont{Thomas}},
  \bibinfo{journal}{Phys. Rev. Lett.} \textbf{\bibinfo{volume}{92}},
  \bibinfo{pages}{150402} (\bibinfo{year}{2004}).

\bibitem{chinSCIENCE2004}
\bibinfo{author}{\bibfnamefont{C.}~\bibnamefont{Chin}},
  \bibinfo{author}{\bibfnamefont{M.}~\bibnamefont{Bartenstein}},
  \bibinfo{author}{\bibfnamefont{A.}~\bibnamefont{Altmeyer}},
  \bibinfo{author}{\bibfnamefont{S.}~\bibnamefont{Riedl}},
  \bibinfo{author}{\bibfnamefont{S.}~\bibnamefont{Jochim}},
  \bibinfo{author}{\bibfnamefont{J.~H.} \bibnamefont{Denschlag}},
  \bibnamefont{and} \bibinfo{author}{\bibfnamefont{R.}~\bibnamefont{Grimm}},
  \bibinfo{journal}{Science} \textbf{\bibinfo{volume}{305}},
  \bibinfo{pages}{1128} (\bibinfo{year}{2004}).

\bibitem{blochrmp08}
\bibinfo{author}{\bibfnamefont{I.}~\bibnamefont{Bloch}},
  \bibinfo{author}{\bibfnamefont{J.}~\bibnamefont{Dalibard}}, \bibnamefont{and}
  \bibinfo{author}{\bibfnamefont{W.}~\bibnamefont{Zwerger}},
  \bibinfo{journal}{Rev. Mod. Phys.} \textbf{\bibinfo{volume}{80}},
  \bibinfo{eid}{885} (\bibinfo{year}{2008}).

\bibitem{pethickbook}
\bibinfo{author}{\bibfnamefont{C.}~\bibnamefont{Pethick}} \bibnamefont{and}
  \bibinfo{author}{\bibfnamefont{H.}~\bibnamefont{Smith}},
  \emph{\bibinfo{title}{Bose-Einstein Condensation in Dilute Gases}}, 
  2nd ed.\ (\bibinfo{publisher}{Cambridge University Press},
  \bibinfo{address}{Cambridge}, \bibinfo{year}{2008}).

\bibitem{Kokkelmans:2002gm}
\bibinfo{author}{\bibfnamefont{S.~J. J. M.~F.} \bibnamefont{Kokkelmans}},
  \bibinfo{author}{\bibfnamefont{J.~N.} \bibnamefont{Milstein}},
  \bibinfo{author}{\bibfnamefont{M.~L.} \bibnamefont{Chiofalo}},
  \bibinfo{author}{\bibfnamefont{R.}~\bibnamefont{Walser}}, \bibnamefont{and}
  \bibinfo{author}{\bibfnamefont{M.~J.} \bibnamefont{Holland}},
  \bibinfo{journal}{Phys. Rev. A} \textbf{\bibinfo{volume}{65}},
  \bibinfo{pages}{053617} (\bibinfo{year}{2002}).

\bibitem{Bruun:2004kz}
\bibinfo{author}{\bibfnamefont{G.~M.} \bibnamefont{Bruun}} \bibnamefont{and}
  \bibinfo{author}{\bibfnamefont{C.~J.} \bibnamefont{Pethick}},
  \bibinfo{journal}{Phys. Rev. Lett.} \textbf{\bibinfo{volume}{92}},
  \bibinfo{pages}{140404} (\bibinfo{year}{2004}).

\bibitem{Dulieu:2009ei}
\bibinfo{author}{\bibfnamefont{O.}~\bibnamefont{Dulieu}} \bibnamefont{and}
  \bibinfo{author}{\bibfnamefont{C.}~\bibnamefont{Gabbanini}},
  \bibinfo{journal}{Rep. Prog. Phys.} \textbf{\bibinfo{volume}{72}},
  \bibinfo{pages}{086401} (\bibinfo{year}{2009}).

\bibitem{Stan:2004ic}
\bibinfo{author}{\bibfnamefont{C.~A.} \bibnamefont{Stan}},
  \bibinfo{author}{\bibfnamefont{M.~W.} \bibnamefont{Zwierlein}},
  \bibinfo{author}{\bibfnamefont{C.~H.} \bibnamefont{Schunck}},
  \bibinfo{author}{\bibfnamefont{S.~M.~F.} \bibnamefont{Raupach}},
  \bibnamefont{and} \bibinfo{author}{\bibfnamefont{W.}~\bibnamefont{Ketterle}},
  \bibinfo{journal}{Phys. Rev. Lett.} \textbf{\bibinfo{volume}{93}},
  \bibinfo{pages}{143001} (\bibinfo{year}{2004}).

\bibitem{Inouye:2004en}
\bibinfo{author}{\bibfnamefont{S.}~\bibnamefont{Inouye}},
  \bibinfo{author}{\bibfnamefont{J.}~\bibnamefont{Goldwin}},
  \bibinfo{author}{\bibfnamefont{M.~L.} \bibnamefont{Olsen}},
  \bibinfo{author}{\bibfnamefont{C.}~\bibnamefont{Ticknor}},
  \bibinfo{author}{\bibfnamefont{J.~L.} \bibnamefont{Bohn}}, \bibnamefont{and}
  \bibinfo{author}{\bibfnamefont{D.~S.} \bibnamefont{Jin}},
  \bibinfo{journal}{Phys. Rev. Lett.} \textbf{\bibinfo{volume}{93}},
  \bibinfo{pages}{183201} (\bibinfo{year}{2004}).

\bibitem{Yabu2003}
\bibinfo{author}{\bibfnamefont{H.}~\bibnamefont{Yabu}},
  \bibinfo{author}{\bibfnamefont{Y.}~\bibnamefont{Takayama}}, \bibnamefont{and}
  \bibinfo{author}{\bibfnamefont{T.}~\bibnamefont{Suzuki}},
  \bibinfo{journal}{Physica B} \textbf{\bibinfo{volume}{329-333}},
  \bibinfo{pages}{25} (\bibinfo{year}{2003}).

\bibitem{Yabu:2004ha}
\bibinfo{author}{\bibfnamefont{H.}~\bibnamefont{Yabu}},
  \bibinfo{author}{\bibfnamefont{Y.}~\bibnamefont{Takayama}},
  \bibinfo{author}{\bibfnamefont{T.}~\bibnamefont{Suzuki}}, \bibnamefont{and}
  \bibinfo{author}{\bibfnamefont{P.}~\bibnamefont{Schuck}},
  \bibinfo{journal}{Nucl. Phys. A} \textbf{\bibinfo{volume}{738}},
  \bibinfo{pages}{273} (\bibinfo{year}{2004}).

\bibitem{adhikariPRA04}
\bibinfo{author}{\bibfnamefont{S.~K.} \bibnamefont{Adhikari}},
  \bibinfo{journal}{Phys. Rev. A} \textbf{\bibinfo{volume}{70}},
  \bibinfo{pages}{043617} (\bibinfo{year}{2004}).

\bibitem{powellPRB05}
\bibinfo{author}{\bibfnamefont{S.}~\bibnamefont{Powell}},
  \bibinfo{author}{\bibfnamefont{S.}~\bibnamefont{Sachdev}}, \bibnamefont{and}
  \bibinfo{author}{\bibfnamefont{H.~P.} \bibnamefont{B\"uchler}},
  \bibinfo{journal}{Phys. Rev. B} \textbf{\bibinfo{volume}{72}},
  \bibinfo{pages}{024534} (\bibinfo{year}{2005}).

\bibitem{Dickerscheid:2005ey}
\bibinfo{author}{\bibfnamefont{D.~B.~M.} \bibnamefont{Dickerscheid}},
  \bibinfo{author}{\bibfnamefont{D.}~\bibnamefont{van Oosten}},
  \bibinfo{author}{\bibfnamefont{E.~J.} \bibnamefont{Tillema}},
  \bibnamefont{and} \bibinfo{author}{\bibfnamefont{H.~T.~C.}
  \bibnamefont{Stoof}}, \bibinfo{journal}{Phys. Rev. Lett.}
  \textbf{\bibinfo{volume}{94}}, \bibinfo{pages}{230404}
  (\bibinfo{year}{2005}).

\bibitem{Zhang:2005dr}
\bibinfo{author}{\bibfnamefont{J.}~\bibnamefont{Zhang}} \bibnamefont{and}
  \bibinfo{author}{\bibfnamefont{H.}~\bibnamefont{Zhai}},
  \bibinfo{journal}{Phys. Rev. A} \textbf{\bibinfo{volume}{72}},
  \bibinfo{pages}{041602(R)} (\bibinfo{year}{2005}).

\bibitem{Avdeenkov:2006ef}
\bibinfo{author}{\bibfnamefont{A.~V.} \bibnamefont{Avdeenkov}},
  \bibinfo{author}{\bibfnamefont{D.~C.~E.} \bibnamefont{Bortolotti}},
  \bibnamefont{and} \bibinfo{author}{\bibfnamefont{J.~L.} \bibnamefont{Bohn}},
  \bibinfo{journal}{Phys. Rev. A} \textbf{\bibinfo{volume}{74}},
  \bibinfo{pages}{012709} (\bibinfo{year}{2006}).

\bibitem{Bortolotti:2008in}
\bibinfo{author}{\bibfnamefont{D.~C.~E.} \bibnamefont{Bortolotti}},
  \bibinfo{author}{\bibfnamefont{A.~V.} \bibnamefont{Avdeenkov}},
  \bibnamefont{and} \bibinfo{author}{\bibfnamefont{J.~L.} \bibnamefont{Bohn}},
  \bibinfo{journal}{Phys. Rev. A} \textbf{\bibinfo{volume}{78}},
  \bibinfo{eid}{063612} (\bibinfo{year}{2008}).

\bibitem{bfrpa}
\bibinfo{author}{\bibfnamefont{S.}~\bibnamefont{Akhanjee}},
  \bibinfo{journal}{Phys. Rev. B} \textbf{\bibinfo{volume}{82}},
  \bibinfo{pages}{075138} (\bibinfo{year}{2010}).

\bibitem{Cazalilla:2004dd}
\bibinfo{author}{\bibfnamefont{M.~A.} \bibnamefont{Cazalilla}},
  \bibinfo{journal}{J. Phys. B: At. Mol. Opt. Phys.}
  \textbf{\bibinfo{volume}{37}}, \bibinfo{pages}{S1} (\bibinfo{year}{2004}).

\bibitem{Cazalilla_review}
\bibinfo{author}{\bibnamefont{\bibnamefont{For a recent review, see}}},
  \bibinfo{author}{\bibfnamefont{M.~A.} \bibnamefont{Cazalilla}},
  \bibinfo{author}{\bibfnamefont{R.}~\bibnamefont{Citro}},
  \bibinfo{author}{\bibfnamefont{T.}~\bibnamefont{Giamarchi}},
  \bibinfo{author}{\bibfnamefont{E.}~\bibnamefont{Orignac}}, \bibnamefont{and}
  \bibinfo{author}{\bibfnamefont{M.}~\bibnamefont{Rigol}},
  \bibinfo{journal}{Rev. Mod. Phys.} \textbf{\bibinfo{volume}{83}},
  \bibinfo{pages}{1405} (\bibinfo{year}{2011}).

\bibitem{cazalillaPRL03}
\bibinfo{author}{\bibfnamefont{M.~A.} \bibnamefont{Cazalilla}}
  \bibnamefont{and} \bibinfo{author}{\bibfnamefont{A.~F.} \bibnamefont{Ho}},
  \bibinfo{journal}{Phys. Rev. Lett.} \textbf{\bibinfo{volume}{91}},
  \bibinfo{pages}{150403} (\bibinfo{year}{2003}).

\bibitem{matheyPRL2004}
\bibinfo{author}{\bibfnamefont{L.}~\bibnamefont{Mathey}},
  \bibinfo{author}{\bibfnamefont{D.-W.} \bibnamefont{Wang}},
  \bibinfo{author}{\bibfnamefont{W.}~\bibnamefont{Hofstetter}},
  \bibinfo{author}{\bibfnamefont{M.~D.} \bibnamefont{Lukin}}, \bibnamefont{and}
  \bibinfo{author}{\bibfnamefont{E.}~\bibnamefont{Demler}},
  \bibinfo{journal}{Phys. Rev. Lett.} \textbf{\bibinfo{volume}{93}},
  \bibinfo{pages}{120404} (\bibinfo{year}{2004}).

\bibitem{matheyPRA2007}
\bibinfo{author}{\bibfnamefont{L.}~\bibnamefont{Mathey}} \bibnamefont{and}
  \bibinfo{author}{\bibfnamefont{D.-W.} \bibnamefont{Wang}},
  \bibinfo{journal}{Phys. Rev. A} \textbf{\bibinfo{volume}{75}},
  \bibinfo{pages}{013612} (\bibinfo{year}{2007}).

\bibitem{Danshita:2013hw}
\bibinfo{author}{\bibfnamefont{I.}~\bibnamefont{Danshita}} \bibnamefont{and}
  \bibinfo{author}{\bibfnamefont{L.}~\bibnamefont{Mathey}},
  \bibinfo{journal}{Phys. Rev. A} \textbf{\bibinfo{volume}{87}},
  \bibinfo{pages}{021603(R)} (\bibinfo{year}{2013}).

\bibitem{tsuchiizuPRA2010}
\bibinfo{author}{\bibfnamefont{E.}~\bibnamefont{Orignac}},
  \bibinfo{author}{\bibfnamefont{M.}~\bibnamefont{Tsuchiizu}},
  \bibnamefont{and} \bibinfo{author}{\bibfnamefont{Y.}~\bibnamefont{Suzumura}},
  \bibinfo{journal}{Phys. Rev. A} \textbf{\bibinfo{volume}{81}},
  \bibinfo{pages}{053626} (\bibinfo{year}{2010}).

\bibitem{Sheehy2005}
\bibinfo{author}{\bibfnamefont{D.~E.} \bibnamefont{Sheehy}} \bibnamefont{and}
  \bibinfo{author}{\bibfnamefont{L.}~\bibnamefont{Radzihovsky}},
  \bibinfo{journal}{Phys. Rev. Lett.} \textbf{\bibinfo{volume}{95}},
  \bibinfo{pages}{130401} (\bibinfo{year}{2005}).

\bibitem{Orignac:2006cb}
\bibinfo{author}{\bibfnamefont{E.}~\bibnamefont{Orignac}} \bibnamefont{and}
  \bibinfo{author}{\bibfnamefont{R.}~\bibnamefont{Citro}},
  \bibinfo{journal}{Phys. Rev. A} \textbf{\bibinfo{volume}{73}},
  \bibinfo{pages}{063611} (\bibinfo{year}{2006}).

\bibitem{dasPRL2003}
\bibinfo{author}{\bibfnamefont{K.~K.} \bibnamefont{Das}},
  \bibinfo{journal}{Phys. Rev. Lett.} \textbf{\bibinfo{volume}{90}},
  \bibinfo{pages}{170403} (\bibinfo{year}{2003}).

\bibitem{polletPRL2006}
\bibinfo{author}{\bibfnamefont{L.}~\bibnamefont{Pollet}},
  \bibinfo{author}{\bibfnamefont{M.}~\bibnamefont{Troyer}},
  \bibinfo{author}{\bibfnamefont{K.}~\bibnamefont{Van~Houcke}},
  \bibnamefont{and} \bibinfo{author}{\bibfnamefont{S.~M.~A.}
  \bibnamefont{Rombouts}}, \bibinfo{journal}{Phys. Rev. Lett.}
  \textbf{\bibinfo{volume}{96}}, \bibinfo{pages}{190402}
  (\bibinfo{year}{2006}).

\bibitem{giambook}
\bibinfo{author}{\bibfnamefont{T.}~\bibnamefont{Giamarchi}},
  \emph{\bibinfo{title}{Quantum Physics in One Dimension}}
  (\bibinfo{publisher}{Oxford University Press}, \bibinfo{address}{Oxford},
  \bibinfo{year}{2003}).

\bibitem{Marchetti2009}
\bibinfo{author}{\bibfnamefont{F.~M.} \bibnamefont{Marchetti}},
  \bibinfo{author}{\bibfnamefont{T.}~\bibnamefont{Jolicoeur}},
  \bibnamefont{and} \bibinfo{author}{\bibfnamefont{M.~M.}
  \bibnamefont{Parish}}, \bibinfo{journal}{Phys. Rev. Lett.}
  \textbf{\bibinfo{volume}{103}}, \bibinfo{pages}{105304}
  (\bibinfo{year}{2009}).

\bibitem{nersesyanPLA1993}
\bibinfo{author}{\bibfnamefont{A.~A.} \bibnamefont{Nersesyan}},
  \bibinfo{author}{\bibfnamefont{A.}~\bibnamefont{Luther}}, \bibnamefont{and}
  \bibinfo{author}{\bibfnamefont{F.~V.} \bibnamefont{Kusmartsev}},
  \bibinfo{journal}{Phys. Lett. A} \textbf{\bibinfo{volume}{176}},
  \bibinfo{pages}{363 } (\bibinfo{year}{1993}).

\bibitem{gogolinbook}
\bibinfo{author}{\bibfnamefont{A.~O.} \bibnamefont{Gogolin}},
  \bibinfo{author}{\bibfnamefont{A.~A.} \bibnamefont{Nersesyan}},
  \bibnamefont{and} \bibinfo{author}{\bibfnamefont{A.~M.}
  \bibnamefont{Tsvelik}}, \emph{\bibinfo{title}{Bosonization and Strongly
  Correlated Systems}} (\bibinfo{publisher}{Cambridge University Press},
  \bibinfo{address}{Cambridge}, \bibinfo{year}{1998}).

\bibitem{yakovenkoJETP1992}
\bibinfo{author}{\bibfnamefont{V.~M.} \bibnamefont{{Yakovenko}}},
  \bibinfo{journal}{JETP Lett.} \textbf{\bibinfo{volume}{56}},
  \bibinfo{pages}{510} (\bibinfo{year}{1992}).

\bibitem{Orignac:1997vq}
\bibinfo{author}{\bibfnamefont{E.}~\bibnamefont{Orignac}} \bibnamefont{and}
  \bibinfo{author}{\bibfnamefont{T.}~\bibnamefont{Giamarchi}},
  \bibinfo{journal}{Phys. Rev. B} \textbf{\bibinfo{volume}{56}},
  \bibinfo{pages}{7167} (\bibinfo{year}{1997}).

\bibitem{Kogut:1979wg}
\bibinfo{author}{\bibfnamefont{J.}~\bibnamefont{Kogut}}, \bibinfo{journal}{Rev.
  Mod. Phys.} \textbf{\bibinfo{volume}{51}}, \bibinfo{pages}{659}
  (\bibinfo{year}{1979}).

\bibitem{Ohta:1979tg}
\bibinfo{author}{\bibfnamefont{T.}~\bibnamefont{Ohta}} \bibnamefont{and}
  \bibinfo{author}{\bibfnamefont{D.}~\bibnamefont{Jasnow}},
  \bibinfo{journal}{Phys. Rev. B} \textbf{\bibinfo{volume}{20}},
  \bibinfo{pages}{139} (\bibinfo{year}{1979}).

\bibitem{Nozieres:1987kg}
\bibinfo{author}{\bibfnamefont{P.}~\bibnamefont{Nozieres}} \bibnamefont{and}
  \bibinfo{author}{\bibfnamefont{F.}~\bibnamefont{Gallet}},
  \bibinfo{journal}{J. Phys. France} \textbf{\bibinfo{volume}{48}},
  \bibinfo{pages}{353} (\bibinfo{year}{1987}).

\bibitem{Muttalib1986}
\bibinfo{author}{\bibfnamefont{K.~A.} \bibnamefont{Muttalib}} \bibnamefont{and}
  \bibinfo{author}{\bibfnamefont{V.~J.} \bibnamefont{Emery}},
  \bibinfo{journal}{Phys. Rev. Lett.} \textbf{\bibinfo{volume}{57}},
  \bibinfo{pages}{1370} (\bibinfo{year}{1986}).

\bibitem{Ledermann:2000wo}
\bibinfo{author}{\bibfnamefont{U.}~\bibnamefont{Ledermann}} \bibnamefont{and}
  \bibinfo{author}{\bibfnamefont{K.}~\bibnamefont{Le~Hur}},
  \bibinfo{journal}{Phys. Rev. B} \textbf{\bibinfo{volume}{61}},
  \bibinfo{pages}{2497} (\bibinfo{year}{2000}).

\bibitem{Folman:2008tx}
\bibinfo{author}{\bibfnamefont{R.}~\bibnamefont{Folman}},
  \bibinfo{author}{\bibfnamefont{P.}~\bibnamefont{Kruger}},
  \bibinfo{author}{\bibfnamefont{J.}~\bibnamefont{Schmiedmayer}},
  \bibinfo{author}{\bibfnamefont{J.}~\bibnamefont{Denschlag}},
  \bibnamefont{and} \bibinfo{author}{\bibfnamefont{C.}~\bibnamefont{Henkel}},
  \bibinfo{journal}{Adv. At. Mol. Opt. Phys.} \textbf{\bibinfo{volume}{48}},
  \bibinfo{pages}{263} (\bibinfo{year}{2002}).

\bibitem{Hofferberth:2008cw}
\bibinfo{author}{\bibfnamefont{S.}~\bibnamefont{Hofferberth}},
  \bibinfo{author}{\bibfnamefont{I.}~\bibnamefont{Lesanovsky}},
  \bibinfo{author}{\bibfnamefont{T.}~\bibnamefont{Schumm}},
  \bibinfo{author}{\bibfnamefont{A.}~\bibnamefont{Imambekov}},
  \bibinfo{author}{\bibfnamefont{V.}~\bibnamefont{Gritsev}},
  \bibinfo{author}{\bibfnamefont{E.}~\bibnamefont{Demler}}, \bibnamefont{and}
  \bibinfo{author}{\bibfnamefont{J.}~\bibnamefont{Schmiedmayer}},
  \bibinfo{journal}{Nat. Phys.} \textbf{\bibinfo{volume}{4}},
  \bibinfo{pages}{489} (\bibinfo{year}{2008}).

\bibitem{Mathey:2008bb}
\bibinfo{author}{\bibfnamefont{L.}~\bibnamefont{Mathey}},
  \bibinfo{author}{\bibfnamefont{E.}~\bibnamefont{Altman}}, \bibnamefont{and}
  \bibinfo{author}{\bibfnamefont{A.}~\bibnamefont{Vishwanath}},
  \bibinfo{journal}{Phys. Rev. Lett.} \textbf{\bibinfo{volume}{100}},
  \bibinfo{pages}{240401} (\bibinfo{year}{2008}).

\bibitem{Molmer:1998wd}
\bibinfo{author}{\bibfnamefont{K.}~\bibnamefont{M{\o}lmer}},
  \bibinfo{journal}{Phys. Rev. Lett.} \textbf{\bibinfo{volume}{80}},
  \bibinfo{pages}{1804} (\bibinfo{year}{1998}).

\bibitem{Recati:2003}
\bibinfo{author}{\bibfnamefont{A.}~\bibnamefont{Recati}},
  \bibinfo{author}{\bibfnamefont{P.~O.} \bibnamefont{Fedichev}},
  \bibinfo{author}{\bibfnamefont{W.}~\bibnamefont{Zwerger}}, \bibnamefont{and}
  \bibinfo{author}{\bibfnamefont{P.}~\bibnamefont{Zoller}},
  \bibinfo{journal}{Phys. Rev. Lett.} \textbf{\bibinfo{volume}{90}},
  \bibinfo{pages}{020401} (\bibinfo{year}{2003}{\natexlab{a}});
  \bibinfo{journal}{J. Opt. B}
  \textbf{\bibinfo{volume}{5}}, \bibinfo{pages}{S55}
  (\bibinfo{year}{2003}{\natexlab{b}}).

\bibitem{Batrouni:2002fs}
\bibinfo{author}{\bibfnamefont{G.~G.} \bibnamefont{Batrouni}},
  \bibinfo{author}{\bibfnamefont{V.}~\bibnamefont{Rousseau}},
  \bibinfo{author}{\bibfnamefont{R.~T.} \bibnamefont{Scalettar}},
  \bibinfo{author}{\bibfnamefont{M.}~\bibnamefont{Rigol}},
  \bibinfo{author}{\bibfnamefont{A.}~\bibnamefont{Muramatsu}},
  \bibinfo{author}{\bibfnamefont{P.~J.~H.} \bibnamefont{Denteneer}},
  \bibnamefont{and} \bibinfo{author}{\bibfnamefont{M.}~\bibnamefont{Troyer}},
  \bibinfo{journal}{Phys. Rev. Lett.} \textbf{\bibinfo{volume}{89}},
  \bibinfo{pages}{117203} (\bibinfo{year}{2002}).

\bibitem{Kollath:2004jc}
\bibinfo{author}{\bibfnamefont{C.}~\bibnamefont{Kollath}},
  \bibinfo{author}{\bibfnamefont{U.}~\bibnamefont{Schollw{\"o}ck}},
  \bibinfo{author}{\bibfnamefont{J.}~\bibnamefont{von Delft}},
  \bibnamefont{and} \bibinfo{author}{\bibfnamefont{W.}~\bibnamefont{Zwerger}},
  \bibinfo{journal}{Phys. Rev. A} \textbf{\bibinfo{volume}{69}},
  \bibinfo{pages}{031601(R)} (\bibinfo{year}{2004}).

\bibitem{Folling:2006dn}
\bibinfo{author}{\bibfnamefont{S.}~\bibnamefont{F{\"o}lling}},
  \bibinfo{author}{\bibfnamefont{A.}~\bibnamefont{Widera}},
  \bibinfo{author}{\bibfnamefont{T.}~\bibnamefont{M{\"u}ller}},
  \bibinfo{author}{\bibfnamefont{F.}~\bibnamefont{Gerbier}}, \bibnamefont{and}
  \bibinfo{author}{\bibfnamefont{I.}~\bibnamefont{Bloch}},
  \bibinfo{journal}{Phys. Rev. Lett.} \textbf{\bibinfo{volume}{97}},
  \bibinfo{pages}{060403} (\bibinfo{year}{2006}).

\end{thebibliography}
\end{document}